\documentclass[sigplan, twocolumn,letterpaper, 10pt]{acmart}
\acmSubmissionID{493}
\renewcommand\footnotetextcopyrightpermission[1]{}
\AtBeginDocument{%
  \providecommand\BibTeX{{%
    \normalfont B\kern-0.5em{\scshape i\kern-0.25em b}\kern-0.8em\TeX}}}

\def\BibTeX{{\rm B\kern-.05em{\sc i\kern-.025em b}\kern-.08em
    T\kern-.1667em\lower.7ex\hbox{E}\kern-.125emX}}

\acmYear{2025}
\copyrightyear{2025}
\setcopyright{acmlicensed}
\acmConference[EuroSys '26]{21th European Conference on Computer Systems}{April 13--April 16, 2026}{Edinburgh, UK}
\acmBooktitle{21th European Conference on Computer Systems (EuroSys '26), April 13--April 16, 2026, Edinburgh, UK}
\acmDOI{XXXXXXX.XXXXXXX}
\acmISBN{978-1-4503-XXXX-X/18/06}

\usepackage{pifont}
\usepackage{newtxtext,newtxmath}
\usepackage{tabularx}
\usepackage{colortbl}
\usepackage{paralist}
\usepackage{mathtools}
\usepackage{multirow}
\usepackage{soul}
\usepackage{graphicx}
\usepackage[linesnumbered,ruled,vlined]{algorithm2e}
\usepackage{amsmath}
\usepackage{wrapfig}
\usepackage{adjustbox}
\usepackage{textcomp}
\usepackage{subcaption}
\usepackage{stfloats}

\newcommand{\sys}{CoCoServe}
\newcommand{\hft}{HFT}
\newcommand{\HFT}{Hugging Face Transformers}
\newcommand{\parabf}[1]{\medskip\noindent\textbf{#1}}

\PassOptionsToPackage{numbers,sort}{natbib}
\usepackage[numbers,sort]{natbib}

\usepackage{enumitem}
\setlist[itemize]{itemsep=0pt, partopsep=0pt, parsep=0pt, topsep=0pt}
\setlist[enumerate]{itemsep=1pt, partopsep=0pt, parsep=0pt, topsep=0pt}

\begin{document}
\title[Unlock the Potential of Fine-grained LLM Serving via Dynamic Module Scaling]{Unlock the Potential of Fine-grained LLM Serving via Dynamic Module Scaling}

\author{Jingfeng Wu$^*$}
\affiliation{%
  \institution{Shenzhen Institutes of Advanced Technology, Chinese Academy of Sciences}
}

\author{Yiyuan He$^*$}
\affiliation{%
  \institution{Shenzhen Institutes of Advanced Technology, Chinese Academy of Sciences}
}

\author{Minxian Xu$^{\dagger}$}
\affiliation{%
  \institution{Shenzhen Institutes of Advanced Technology, Chinese Academy of Sciences}
}

\author{Xitong Gao}
\affiliation{%
  \institution{Shenzhen Institutes of Advanced Technology, Chinese Academy of Sciences}
}

\author{Kejiang Ye}
\affiliation{%
  \institution{Shenzhen Institutes of Advanced Technology, Chinese Academy of Sciences}
}

\author{Chengzhong Xu}
\affiliation{
    \institution{University of Macau}
}

\thanks{
    $^*$Equal Contributions \\
    $^{\dagger}$Minxian Xu is the corresponding author (mx.xu@siat.ac.cn).
}

\begin{abstract}
The rise of large language models (LLMs) has created new opportunities across various fields but has also introduced significant challenges in resource management. Current LLM serving systems face a fundamental tension: balancing serving demands with limited resources while adapting to unpredictable traffic patterns. Static deployments lead to suboptimal resource utilization and performance degradation under dynamic workloads. Furthermore, the high cost of adjusting instances hinders dynamic scaling, limiting the true potential of efficient LLM serving.

To address this, we propose \sys{}, an elastic system that facilitates dynamic and fine-grained scaling. Its key innovation lies in the module-level operations for the replication and migration of LLM modules, such as decoder layers and projections. Through a comprehensive analysis of the trade-offs associated with these operations, we develop an auto-scaling mechanism that dynamically regulates module-level resource allocation and performance optimization, enabling a more cost-effective deployment of LLMs. Our evaluation demonstrates that the scaling operations employed by \sys{} exhibit excellent scalability and can reduce costs by 46\% while maintaining availability. Compared to state-of-the-art LLM serving systems (e.g., \HFT{} and vLLM), our approach reduces latency by 14\%-75\% and achieves 1.16$\times$-4$\times$ throughput on average across different model sizes and workloads.

\end{abstract}

\ccsdesc[500]{Computing methodologies~Artificial intelligence}
\ccsdesc[300]{Computer systems organization~Cloud computing}

\keywords{LLM, Module Scaling, Replication, Migration}

\maketitle 

\vspace{-0.2cm}
\section{Introduction}
\label{sec:intro}
With the emergence of high-performance LLMs such as GPT~\cite{report23-gpt4}, LLaMA~\cite{llama2}, and DeepSeek~\cite{DeepSeekV3TR}, LLM serving has introduced unprecedented user experiences and accelerated the adoption of the Model-as-a-Service (MaaS) paradigm~\cite{MaaS}. In recent years, an increasing number of MaaS products have entered the market, covering a wide range of application scenarios, including conversational AI~\cite{chatbot,chat1}, code assistance~\cite{codeAssistance1,codeAssistance2}, and image generation~\cite{openai-dalle3}.

However, the other side of the story reveals significant challenges in resource utilization. LLMs are well-known for their substantial resource consumption, particularly regarding memory and computational power, which imposes considerable operational pressures and high costs on MaaS providers~\cite{openai-api}. Furthermore, unlike traditional internet services, the online inference process for LLM serving often dominates the end-to-end response time, leading to reduced request throughput~\cite{xiao2024smoothquantaccurateefficientposttraining, zhang2024plugandplay, lin2024awqactivationawareweightquantization, sun2024simpleeffectivepruningapproach}.

\begin{figure}[t]
    \centering
    \includegraphics[width=0.9\linewidth]{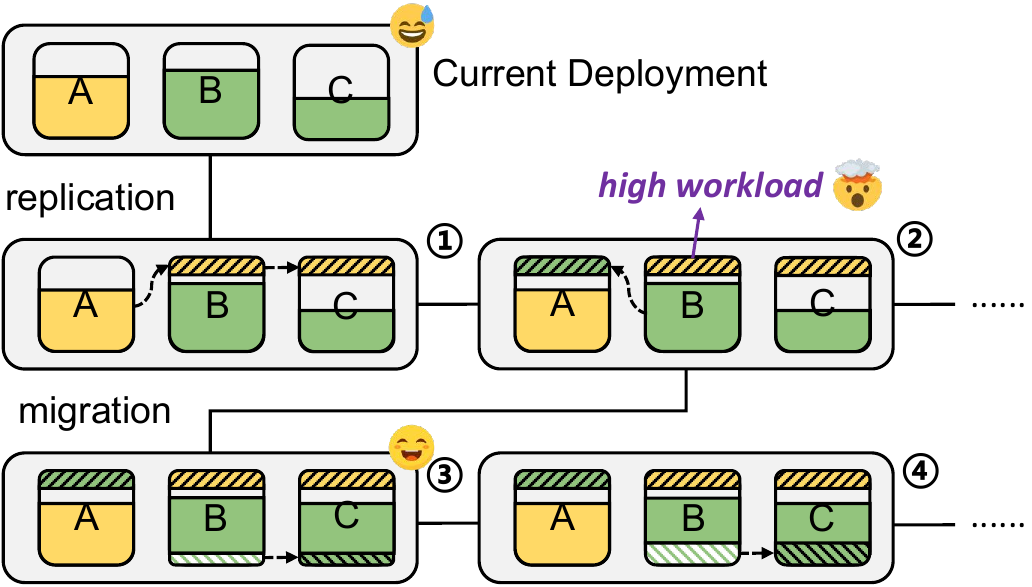}
    \caption{Illustration of the scaling mechanism in \sys{}. Yellow and green blocks represent different instances deployed across three devices labeled A, B, and C. Blocks filled with left-slanted and right-slanted lines represent replicated and migrated modules respectively.}
    \label{fig:intro}
\end{figure}

There are several concerns associated with addressing the aforementioned challenges. First, during low workload periods, resource demands of model instances may fall below the resources provided by the cluster, resulting in idle resources. Second, excessive workloads can lead to performance degradation, violations of Service Level Objectives (SLOs)~\cite{eurosys25-DLWorkloads,wu2025cloudnativesim}, or even out-of-memory (OOM) failures~\cite{alizadeh2024llmflashefficientlarge}. Finally, making instance-level adjustments to adapt to dynamic workloads incurs significant overhead and performance penalties.

To achieve satisfactory performance of LLM serving systems, recent research has focused on dynamic scaling mechanisms. These approaches ~\cite{miao2024spotserve,osdi24-llumnix,choi2022serving,crankshaw2020inferline,pang2025optimizingllminferencethroughput} have made practical attempts at various levels, such as online migration, customized replication strategies for LLM inference phases, and adjustments to resource allocation and batch sizes across different instance replicas. While these approaches offer several advantages, the existing solutions treat the LLM instance as the minimal unit of scaling, attempting to replicate and migrate it to adapt to dynamic environments~\cite{You2025AlloyStackAL, Wang2024LoRAFlowDL}. Although this seems reasonable in multi-instance deployment scenarios, each LLM instance potentially occupies hundreds of gigabytes of GPU memory. Given that models are composed of various modules\footnote{In this paper, the modules refer to decoder layers, attention,  feed-forward network, projections, and key-value cache.}, for example, a LLaMA-13B model consists of 40 decoder layers (i.e., transformer blocks)~\cite{ollama}, implementing dynamic scaling at the module level would enable more precise control over both resources and performance for LLM serving systems.

Based on the above motivation, we propose a novel scaling mechanism leveraging module replication and migration for LLMs. Module replication enables flexible and precise performance enhancement for instances at a controlled cost, effectively utilizing idle resource fragments caused by mismatches between model sizes and allocated resources. Meanwhile, module migration alleviates resource imbalance across devices, preventing catastrophic failures due to computational or memory overload on certain devices. 

Fig. \ref{fig:intro} illustrates the scaling mechanism based on replication and migration operations. The top of the figure depicts an unsatisfactory current deployment with two instances running on three GPUs, where resource under-utilization and load imbalance arise due to misaligned instance sizes and available resources. Our proposed system, \sys{}, searches for an optimal operation strategy: First, by replicating modules, i.e. copying part of the yellow LLM instance to devices B and C, and part of the green LLM instance to device A, which reuses idle resources efficiently. Second, if device B in State 2 experiences excessive load leading to insufficient memory or computational overload, migrating a portion of its modules to device C mitigates the issue. Depending on the scale of migration, States 3 and 4 can be transited, but only State 3 achieves a satisfactory deployment.

By leveraging this module-level scaling mechanism, we have developed \sys{}, an efficient LLM serving system that supports fine-grained scalability. Our evaluation demonstrates that the scaling operations used by \sys{} require only 0.3 seconds to execute, and reduce cost by over 46\% while maintaining availability. Under various requests per second (RPS) test scenarios, our system demonstrates significant performance advantages. When compared to the Hugging Face Transformers (HFT)~\cite{wolf-etal-2020-transformers}, CoCoServe reduces latency by 61\%-75\% and achieves 2.7$\times$-4.0$\times$ throughput improvements on average. Even against another state-of-the-art system vLLM~\cite{sosp23-vllm}, CoCoServe maintains impressive efficiency gains, reducing latency by 14\%-32\% and increasing throughput by 1.16x-1.48x on average across different workloads.

Our key \textbf{contributions} can be summarized as follows:
\begin{itemize}[leftmargin=2em]
    \item We introduce module-level scaling operations for LLM module replication and migration, and conduct detailed analysis of these operations to validate its feasibility.
     \vspace{-0.05cm}
    \item We design an auto-scaling mechanism based on module replication and migration, which maintains performance under fluctuating workloads.
    \vspace{-0.05cm}
    \item We develop \sys{}, an elastic LLM serving system that supports dynamic, fine-grained and module scaling, while  compatible with existing frameworks.
    \vspace{-0.05cm}
    \item We conduct comprehensive experimental evaluation of \sys{}, demonstrating its superior performance compared to state-of-the-art systems \hft{} and vLLM.
\end{itemize}

\section{Background and Motivation}
\label{sec:bg}
In this section, we introduce the background of LLM inference, analyze the limitations of existing solutions and highlight the challenges that remain unresolved, and present the motivation behind our proposed solution.

\subsection{LLM Inference} 
LLM inference serves as the fundamental operation in LLM serving systems, yet it exhibits markedly higher latency than traditional web service search queries. This performance disparity originates from two key factors: (1) the auto-regressive token-generation process (decoding) that requires sequential computation for each output token, and (2) the inherent architectural complexity of modern LLMs, which typically consist of hundreds of interconnected modules. 

For example, the LLaMA-13B model architecture consists of an embedding layer followed by 40 transformer decoder layers. Each decoder layer contains approximately 317 million parameters~\cite{ollama}, distributed across multiple components: self-attention with projections for query (Q), key (K), value (V), and output (O), as well as feed-forward network (FFN) projections. These projections are linear transformations that map input vectors to different representation spaces—the Q/K/V projections transform inputs for attention computation while the O projection converts attention outputs back to the model dimension. Similarly, the FFN projections first expand the representation to a higher dimension to capture more complex patterns, then project it back to the original dimension. The Q/K/V/O projections each require $d_{\text{model}}^2$ parameters (where $d_{\text{model}}=5120$), totaling $4 \times d_{\text{model}}^2$ parameters for the attention mechanism. The feed-forward network contains three layers with dimensions $d_{\text{model}} \times d_{\text{ff}}$ and $d_{\text{ff}} \times d_{\text{model}}$ (where $d_{\text{ff}}=13824$), plus additional parameters for layer normalization. 

Auto-regressive decoding necessitates repeated forward passes through the entire model for each new token~\cite{vaswani2017attention}, with dependencies on all previously generated tokens. To mitigate redundant computation, modern inference systems employ key-value (KV) cache~\cite{li2024llminferenceservingsurvey,li2024optimizingmixtureofexpertsinferencetime, qin2024mooncakekvcachecentricdisaggregatedarchitecture, eurosys25-spinfer, zhang2023h2o}, which stores intermediate attention states and avoids reprocessing historical tokens. However, this optimization introduces memory-bandwidth constraints, as the KV cache grows linearly with sequence length, further exacerbating latency. Additionally, based on whether KV cache is generated, the LLM inference phase can be divided into two distinct phases: prefill and decode~\cite{Patel2023SplitwiseEG}. During the prefill phase, the model processes the entire prompt at once, generating initial KV cache entries for all input tokens. This phase is compute-bound and benefits from high GPU utilization through batched operations~\cite{10.5555/3692070.3693589}. The subsequent decode phase generates new tokens one by one, with each token requiring a full forward pass through the model while referencing the growing KV cache. This phase is typically memory-bound due to frequent cache access operations. The prevalent paradigm is based on multi-instance scenario by deploying multiple model replicas across devices (such as GPUs)~\cite{hu2024inferenceinterferencedisaggregatellm}. This architectural approach offers  advantages including stronger scalability, enhanced throughput, and inherent fault tolerance against individual instance failures, while suffers from corse-grained resource management.



\subsection{Limitations of Existing Solutions}
Recent years, researchers have promoted the development of numerous scaling approaches operating at multiple system levels. These approaches target different objects of the serving pipeline, which we classify into three primary categories based on their optimization scope.

\textbf{Workload-level Scaling.} To achieve load balancing across instances and devices, recent work has focused on workload scaling~\cite{stojkovic2024dynamollmdesigningllminference, eurosys25-skyserve,arxiv24-burstgpt} . For example, Llumnix~\cite{osdi24-llumnix} implements cross-instance dynamic migration to balance loads adaptively, while USHER~\cite{osdi24-usher} employs workload division holistically for instance replicas to achieve a balance between computational utilization and storage utilization. Though workload balancing, these methods operate within fixed computational boundaries and do not further adjust the performance of the instances themselves.

\textbf{Stage-level Scaling.} Systems like LoongServe~\cite{Wu2024LoongServeES} and DistServe~\cite{osdi24-distserve} focus on the two phases (prefill and decode) of LLM. They analyze the significant differences between these two phases, which are computationally intensive and memory-intensive, respectively, and tailor distinct scaling strategies for each through parallelization. While optimizations for different stages can achieve good results in specific scenarios, such as long-context processing, they also tend to create resource utilization imbalances, such as prefill lacks memory efficiency, while decode under-utilizes compute resources.

\textbf{Device-level Scaling.} SkyServe~\cite{eurosys25-skyserve} determines the optimal number of replicas across diverse failure domains, such as regions and cloud providers, based on the target workload. AlpaServe~\cite{osdi23-alpaserve} explores a new spectrum of hybrid parallelism strategies, seeking efficient placement and parallelization of large-scale deep learning models within distributed clusters. While device-level scaling places greater emphasis on automated configuration in distributed environments, it often faces challenges due to the substantial overhead associated with instance operations, which can hinder the responsiveness and efficiency of scaling decisions.

The above systems have proposed effective scaling mechanisms at different levels and achieved practical success in their respective scenarios. We observe that two fundamental operations including replication and migration recur across different levels, forming the basis for scaling LLM serving capacity. This is because replication enables load distribution across instances while also providing parallel acceleration benefits. On the other hand, migration allows for adjustments in deployment strategies, better adapting to dynamic environments while improving load balancing across devices. However, existing solutions have the following shortcomings in executing these operations: (a)\textbf{ High overhead in instance-level replication}: Replicating entire instances often involves one or more LLMs, resulting in coarse granularity that hinders dynamic scheduling. (b) \textbf{Mismatch between instance sizes and allocated resources}: The disparity frequently leads to fragmented idle resources, reducing overall utilization efficiency. (c) \textbf{Imbalance between Devices}: While request migration balances load across instances, device-level imbalances persist, with some devices remaining memory-bound while others become compute-bound.

\subsection{Challenges and Motivation}
However, when meeting dynamic workload patterns induced by fluctuating request, the multi-instance paradigm introduces complex resource allocation challenges: 

\begin{figure}[htbp]
    \centering
    \includegraphics[width=0.82\linewidth]{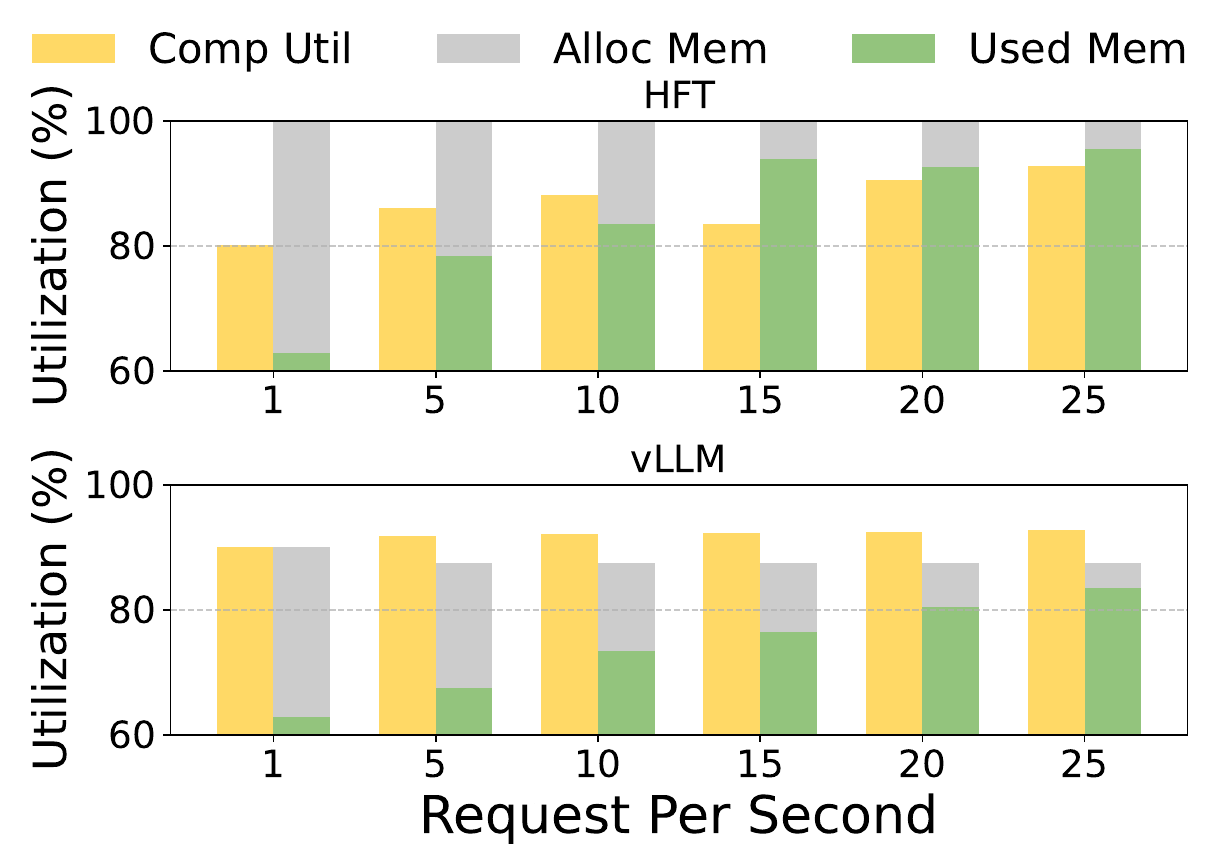}
    \caption{GPU resource utilization comparison between HFT and vLLM serving frameworks across different request rates, conducted with a single LLaMA-13B instance deployed on an A100 GPU, with each test repeated five times.}
    \label{fig:bg1}
\end{figure}
\textbf{Suboptimal Utilization.} Modern LLM serving systems often suffer from suboptimal resource utilization due to two primary factors. First, at low request rates (RPS $\leq$ 10), significant computational and memory resources remain idle due to static resource allocation, as illustrated in Fig. \ref{fig:bg1}. The data illustrates that Hugging Face Transformers (HFT) and vLLM implementations leave approximately 20\%-40\% of GPU resources unused under these conditions. utilization inefficiency. Furthermore, the mismatch between the given resource constraints and the dynamic demands of LLM serving also leads to resource utilization inefficiency. Traditional systems allocate fixed resources, which often means that satisfying storage requirements makes it difficult to simultaneously meet computational needs, or vice versa. This allocation-demand mismatch results motivates us to achieve dynamic resource allocation to improve the overall cost-effectiveness of the deployment.

\begin{figure}[htbp]
    \centering
    \includegraphics[width=0.75\linewidth]{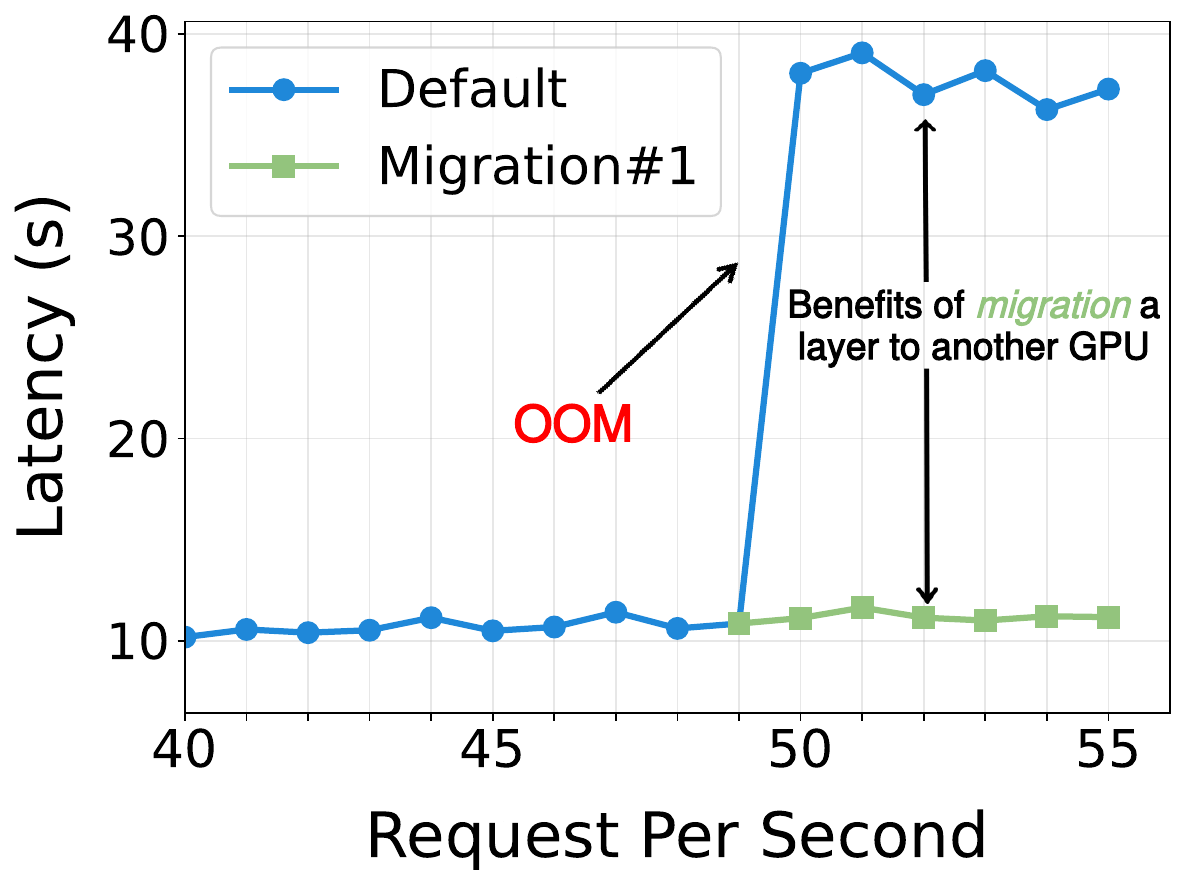}
    \caption{Performance comparison between the default configuration and migrating 1 layer to another device under high load conditions, conducted with a single LLM-13B instance deployed on an A100 GPU across varying RPS rates.}
    \label{fig:bg2}
\end{figure}

%
\textbf{SLO Violation.} LLM serving systems frequently encounter performance degradation under high load conditions, resulting in SLO violations~\cite{10.1145/3419111.3421284,273804} and potentially catastrophic OOM failures~\cite{alizadeh2024llmflashefficientlarge}. As request rates increase, the system's ability to maintain consistent response times deteriorates significantly. As illustrated in Fig. \ref{fig:bg2}, when the RPS rate exceeds 50, the default configuration experiences a dramatic latency increase, reaching up to 37s. This performance cliff effect is caused by OOM failures, leads to SLO violation rates escalating rapidly, compromising service reliability and user experience.

While adjusting batch sizes can temporarily mitigate these issues, this approach suffers from request blocking and reduced overall throughput. This motivates us that strategic layer migration offers a more effective solution without degrading instance performance. As demonstrated in Fig. \ref{fig:bg2}, by migrating only 1 decoder layer to another GPU (Migration \#1), the system maintains latency around 11.2 seconds even at high request rates (50-55 RPS), representing a substantial 70\% reduction in latency compared to the default configuration. This migration approach preserves throughput while ensuring consistent performance, effectively preventing SLO violations and OOM failures.

\textbf{Costly Instance Adjustments.}
Runtime adjustment of model instances, such as replication and migration, introduces substantial operational overhead. Our observations show that performing a single adjustment for a 13B model takes 8–25 seconds and consumes an additional 20–30 GB of GPU memory. These adjustments not only incur significant latency penalties but also create resource contention, which can degrade overall system performance during the transition period. Traditional instance-level operations often require complete model reloading or extensive state transfers, resulting in service disruptions that can last from several seconds to even minutes. This motivates us to use module-level operations that offer a more fine-grained approach to instance adjustment with significantly reduced overhead. By operating at the layer or smaller components level rather than manipulating entire model instances, systems can achieve fine-grained resource allocation that closely matches workload demands. 


\section{Module-level Operations and Analysis}
\label{sec:operation}

To exploit the effects and scope of fine-grained scaling, we implemented simple yet effective replication and migration at the module level and conducted analysis on them. In this section, we first introduce the design and implementation of these two primitive operators, then through experimental analysis of these operations under increasing RPS, we derived their characteristics, which form the foundation for auto-scaling algorithm in our system.
\subsection{Replication and Migration Design}
\label{sec:operation_def}
\begin{figure}[htbp]
    \centering
    \includegraphics[width=0.9\linewidth]{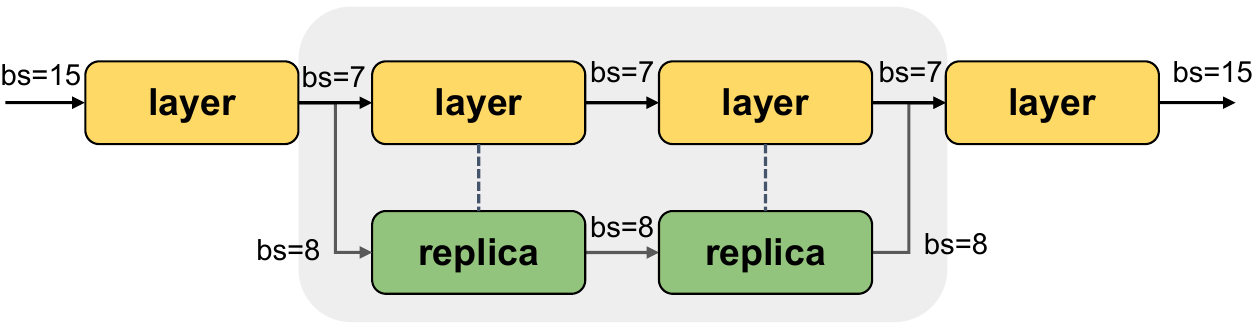}
    \caption{Illustration of the replication for decoder layers. Yellow blocks represent the yellow layers deployed on the main device, while green blocks represent replicas deployed on another device.}
    \label{fig:replication}
\end{figure}
\textbf{Replication.} The replication operation creates replicas of selected model modules across available devices, enabling parallel processing of requests through shared data computation. As shown in Fig. \ref{fig:replication}, the operation reconstructs the computational topology by establishing new dataflow paths between model layers. In the default configuration, each device processes a fixed batch size of 15 requests. After applying replication, the system distributes workloads more precisely across two replicas, processing batches of 7 and 8 requests respectively. In this process, as the system needs to distribute requests to replicas and collect data from replicas, it introduces additional overhead of scatter operation and an all-gather operation. However, for consecutive layers, these additional overheads only appear at their beginning and end points, while internally they will forward like the original layers. Evidently, the continuity between replicas affects the communication overhead of module-level replication. The module-level replication demonstrates particular value in resource-constrained environments through localized data parallelism to accelerate computation and improved memory utilization by aggregating fragmented memory spaces.

\begin{figure}[htbp]
    \centering
    \includegraphics[width=\linewidth]{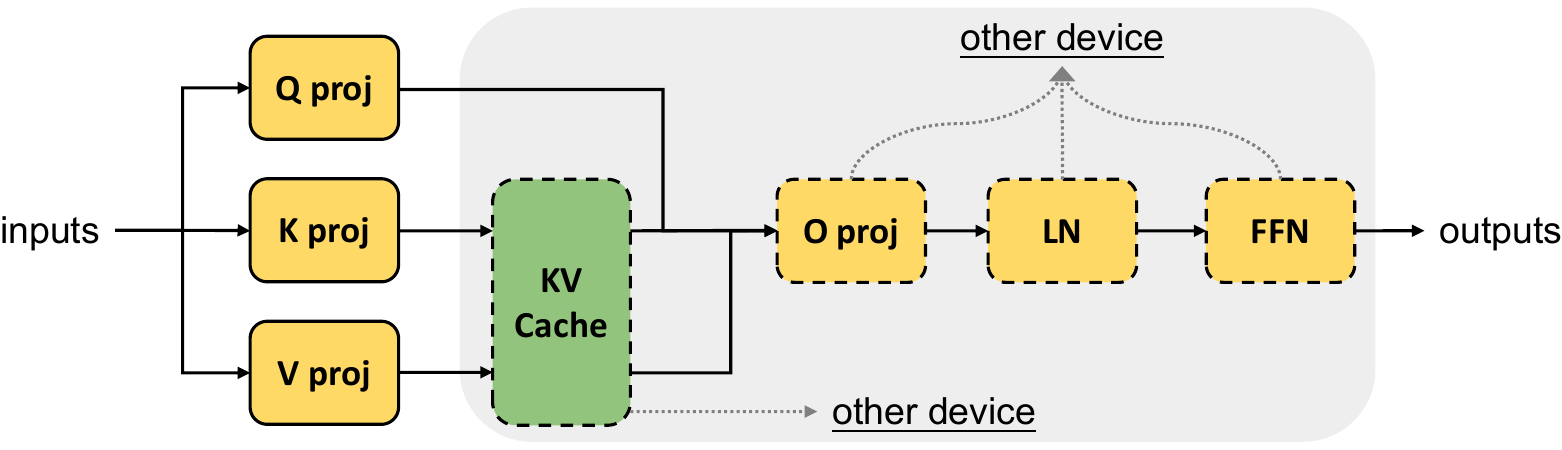}
    \caption{Illustration of the migration for modules in a decoder layer. Yellow modules are computation-intensive, while green modules are memory-intensive. The migration operation allows them to be transferred to different devices respectively.}
    \label{fig:migration}
\end{figure}

\textbf{Migration.} The migration operation enables transferring modules at different granularity levels to other devices for fine-grained deployment adjustments. As shown in Fig. \ref{fig:migration}, it also enables migration of fine-grained components within layers, such as projections and KV caches, which demonstrates that computation-intensive modules (highlighted in yellow) and memory-intensive modules (shown in green) can be migrated to separate devices that better match their resource requirements. This operation allows for the redeployment of modules with different characteristics, offering two core advantages in dynamic load environments: 1) It can mitigate OOM failures by relocating memory-intensive modules or entire layers, while also preventing performance degradation and SLO violations caused by computational overload. 2) It enables more balanced resource utilization by aligning module characteristics with device capabilities through the  placement of modules to avoid uneven utilization of computational and storage resources across devices.


\textbf{Implementation.} We leverage the hook mechanism to implement module replication and migration. Using pre-hooks and post-hooks, we transfer input data and computation results across GPUs, enabling dynamic scaling without modifying the original model architecture. For migration, we replicate the target module—with its weights and caches—to the designated machine and remove the local copy. For replication, the module is duplicated with similar hook-based data transmission. Additionally, when operating on layers, our system allows optional migration of the corresponding KV cache along with the layer.

Furthermore, during replication and migration, it is essential to maintain system correctness throughout execution. This includes ensuring proper input allocation to replicas, accurate aggregation of results, and preservation of model semantics during these operations. Additionally, potential interference between replicated layers and pre-existing modules on the target device must be mitigated, as resource contention may degrade the performance of previously deployed components. To address this, we deliberately avoid monopolizing device resources and strive to preserve sequential consistency among replicas on each device, thereby minimizing interference and memory contention.

\subsection{Observations of Replication Performance}
\label{sec:analysis_replication}

\begin{figure}[h]
    \centering
    \begin{subfigure}[b]{0.48\linewidth}
        \centering
        \includegraphics[width=\linewidth]{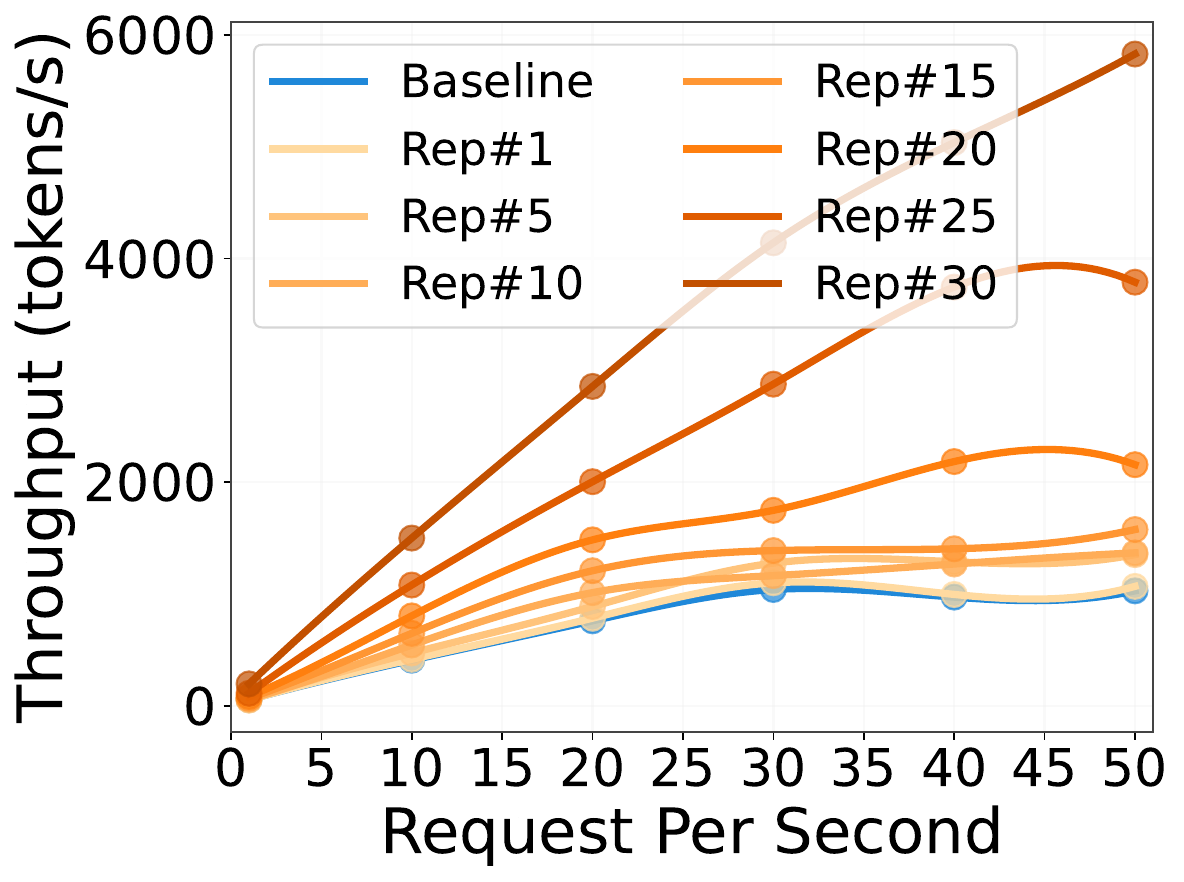}
        \caption{Throughput varying replicas}
        \label{fig:layer_rep_throughput}
    \end{subfigure}
    \hfill
    \begin{subfigure}[b]{0.48\linewidth}
        \centering
        \includegraphics[width=\linewidth]{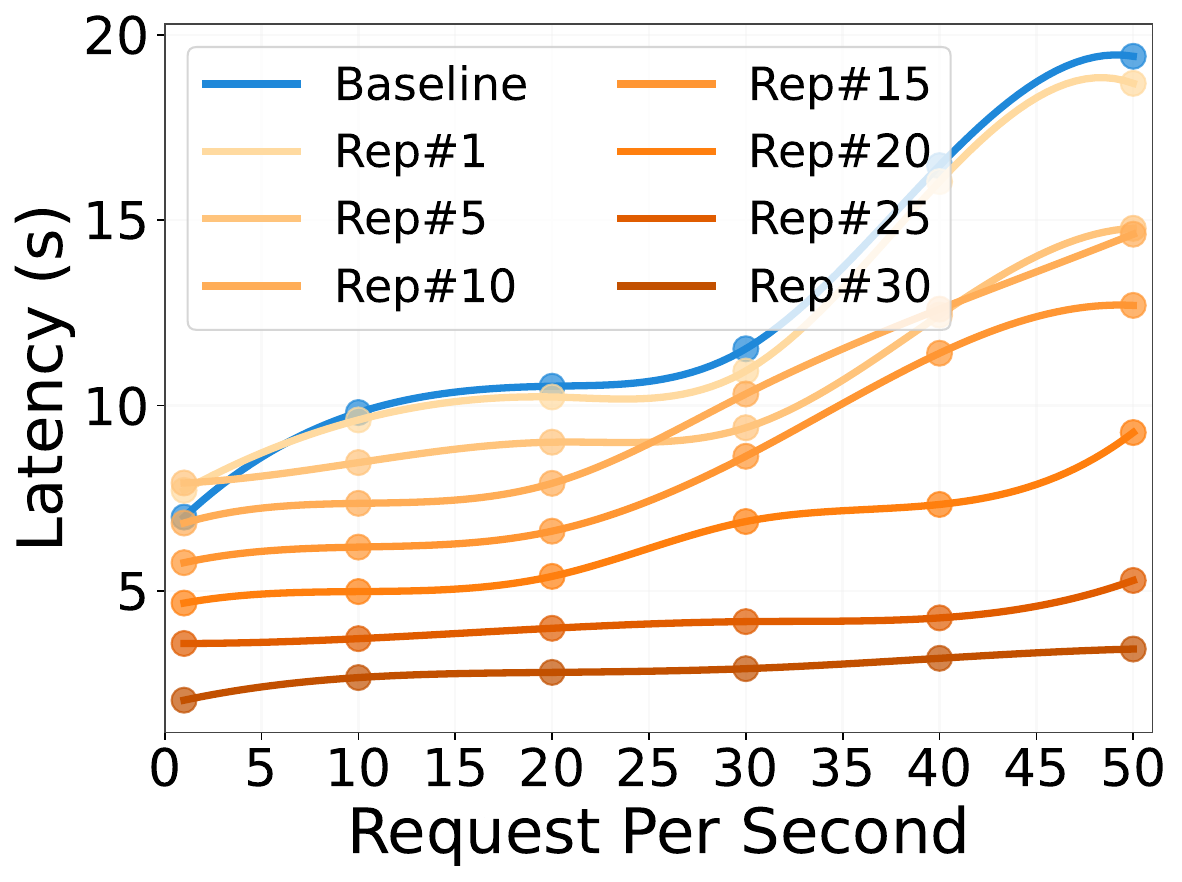}
        \caption{Latency varying replicas}
        \label{fig:layer_rep_latency}
    \end{subfigure}
    
    \begin{subfigure}[b]{0.48\linewidth}
        \centering
        \includegraphics[width=\linewidth]{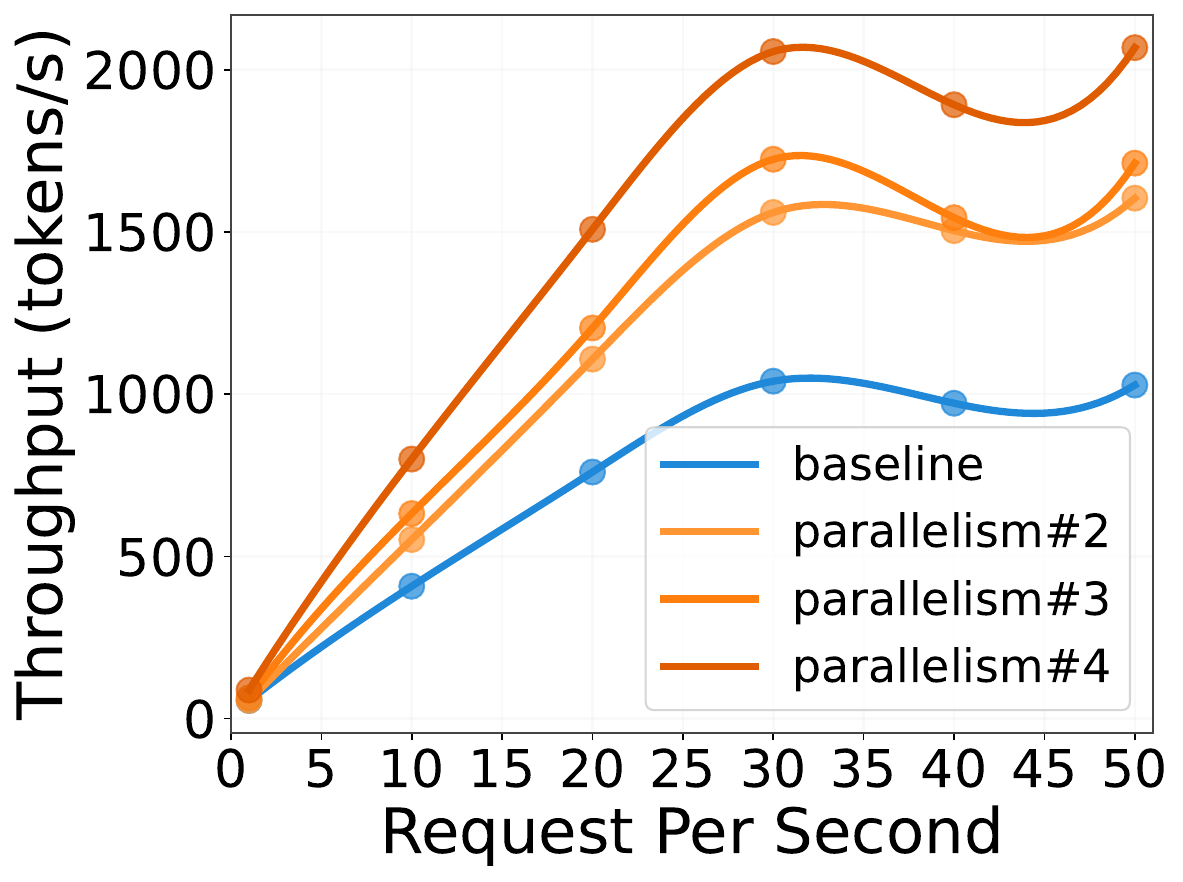}
        \caption{Throughput varying dop}
        \label{fig:parall_throughput}
    \end{subfigure}
    \hfill
    \begin{subfigure}[b]{0.48\linewidth}
        \centering
        \includegraphics[width=\linewidth]{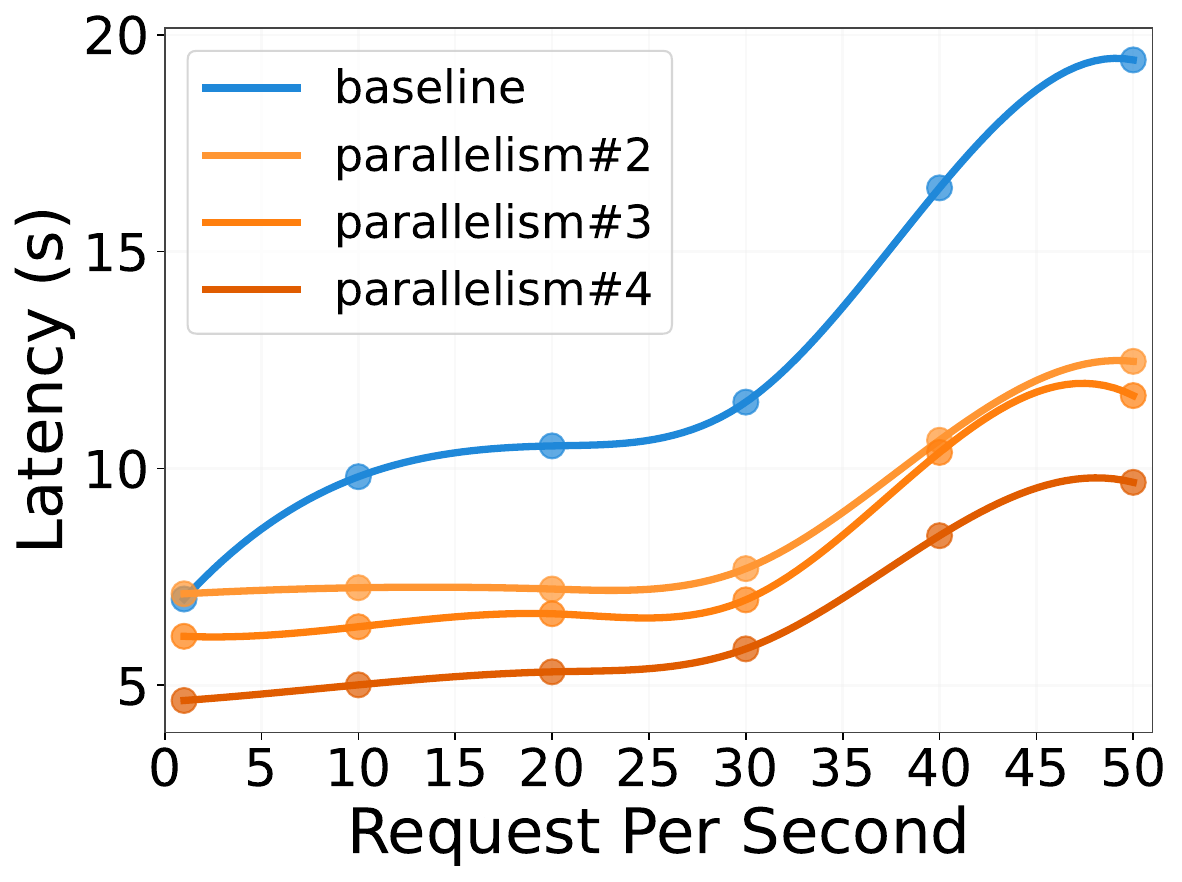}
        \caption{Latency varying dop}
        \label{fig:parall_latency}
    \end{subfigure}
    \caption{Performance analysis of layer replication and parallelism strategies under varying request rates}
    \label{fig:replication_analysis}
\end{figure}

To investigate the performance impact of two key dimensions in module-level replication—namely, layer replication count (the number of replicated layers of the model) and degree of parallelism (dop, indicating the number of concurrent layers)—we conducted experiments using the LLaMA-13B model on 4 NVIDIA A100 GPUs. The baseline configuration represents a completely unmodified serial execution environment without any parallelism or layer replication. Fig. \ref{fig:replication_analysis} presents comprehensive results demonstrating how these factors affect system performance under varying request loads.

In Figs \ref{fig:layer_rep_throughput} and \ref{fig:layer_rep_latency}, we maintain a fixed parallelism degree of 2 while progressively increasing the layer replication count from baseline to 30 layers. The throughput results in Fig. \ref{fig:layer_rep_throughput} reveal that more layer replication substantially enhances system capacity, with performance gains accelerating non-linearly at higher replication counts. At peak load (50 RPS), the 30-layer configuration achieves 329.42\% higher throughput than baseline (4.3× improvement), while 25-layer and 20-layer replications demonstrate 193.10\% and 90.80\% improvements respectively. This progressive enhancement confirms that  layer replication effectively distributes computational load through more replicas.

The corresponding latency measurements in Fig. \ref{fig:layer_rep_latency} exhibit complementary behavior. While the baseline system suffers from exponentially increasing latency (reaching 20s at 50 RPS), configurations with more replicas maintain stable response times. The 30-layer replication shows particular resilience, sustaining sub-5s latency across all load levels - a 75.80\% average improvement over baseline. Notably, even moderate 15-layer replication achieves 45.07\% higher throughput with 30.37\% latency reduction, demonstrating the effectiveness of partial model replication.

Figs \ref{fig:parall_throughput} and \ref{fig:parall_latency} examine parallel processing by fixing replication depth at 20 layers while scaling parallelism from baseline to 4-way configuration. The results reveal two distinct operational regimes: below 30 RPS, quadruple parallelism delivers near-linear scaling (95.26\% throughput increase with 48.76\% latency reduction), while higher loads expose diminishing returns. With 50 RPS, 4-way parallelism achieves 163.68\% throughput improvement versus 268.48\% from equivalent-depth layer replication (Rep\#25), suggesting that deeper model parallelism may be more effective than wider data parallelism for high-intensity workloads. This performance plateau at extreme loads indicates potential resource contention or synchronization overhead in pure data-parallel configurations.

From a holistic perspective, our observations demonstrate three key advantages of layer replication. First, it achieves superior scalability by flexibly adjusting replication depth and parallelism degree, unlike the all-or-nothing requirement of full-model replication. Second, it maintains exceptional stability under heavy loads, delivering consistent sub-5s latency where baseline systems suffer exponential degradation. Finally, layer replication enables partial data-parallel effects that exhibit nonlinear positive correlation with replication count and parallelism degree. This demonstrates that layer replication provides a more resource-efficient pathway for performance scaling, particularly valuable in constrained GPU environments by effectively utilizing idle resource fragments.

\subsection{Observations of Migration Granularity}
\label{sec:analysis_migration}
\begin{table}[htbp]
\centering
\caption{Module Memory and Computation Analysis}
\begin{tabular}{|l|c|c|}
\hline
\textbf{Module} & \textbf{Memory} & \textbf{Computation} \\
\hline
self\_attn.q/k/v/o\_proj & 50 MB & 13.42 GFLOPs \\
\hline
self\_attn & 200 MB & 55.02 GFLOPS \\
\hline
ffn.gate/up/dwon\_proj & 135 MB & 36.24 GFLOPs \\
\hline
decoder layer & 605 MB & 127.5 GFLOPs \\
\hline
\end{tabular}
\label{tb:size}
\end{table}

To explore the feasible granularity of migration operations, we conducted an experimental and theoretical analysis~\cite{Korthikanti2022ReducingAR} on the LLaMA-13B model under standard inference conditions (batch size = 1, sequence length = 256, bfloat16 precision), focusing on weight memory consumption and computational demands while excluding normalization, embedding, and activation variables. The attention module (self\_attn) requires 200 MB of memory for storing weights and generates 55.02 GFLOPS of computational overhead. Within this module, the Q/K/V/O projections, with dimensions [5120, 5120], account for 50 MB of memory and contribute 13.42 GFLOPS from GEMM operations, along with an additional 1.34 GFLOPS for attention score calculations. Meanwhile, the feed-forward network components—comprising gate, up, and down projections with dimensions [5120, 13824]—occupy 135 MB of memory and introduce 36.24 GFLOPS of computational load. When considering a full decoder layer, which integrates these components alongside normalization layers and activation memory, the total memory footprint reaches 605 MB, accompanied by 127.5 GFLOPS of computational demand.

From a computational characteristics perspective, each module exhibits distinct resource utilization patterns. The attention and feed-forward modules are primarily computation-intensive, exhibiting high computational density relative to their memory footprint (0.275 GFLOPs/MB for self-attention and 0.268 GFLOPs/MB for FFN based on the table data). In contrast, the KV cache represents a memory-intensive component, exhibiting dynamic memory consumption during execution that fluctuates between several hundred megabytes to a few gigabytes, with its actual footprint heavily dependent on runtime parameters such as batch size and sequence length. 

Based on our module analysis, we recommend several fine-grained migration strategies. For SLO violations and OOM failures under high workloads, migrating entire layers when possible reduces communication overhead while maintaining effectiveness. For devices with computational capacity exceeding memory, migrating computation-intensive modules such as attention projections and feed-forward networks optimally leverages available computational resources. Conversely, for devices with memory capacity exceeding computational power, migrating the KV Cache proves advantageous as it requires significant memory but minimal computation, with its dynamic memory footprint enabling flexible adaptation to available resources. This fine-grained approach at the layer level enables precise resource management aligned with each module's distinct characteristics.

\section{Dynamic Auto-Scaling Mechanism Design}
\label{sec:mechanism}
To address the challenges arising from static and coarse-grained resource usage, we design the auto-scaling mechanism of \sys{} that leverages module-level operations based on Section \ref{sec:operation} to scale-up and scale-down instances. 

\vspace{-0.3cm}
\subsection{Scale-Up: Speedup via Layer Replication}

To achieve efficient scale-up during low workload periods, we employ a novel approach that leverages parallelism through layer-level replication to enhance computational performance. In contrast to traditional instance-based scale-up methods,  our design achieves lower cost by re-utilize the idle resources fragments rather than requiring additional capacity. However, as previously analyzed, layer-level replication introduces significant combinatorial complexity, with exhaustive enumeration of all possible layer permutations across expandable nodes yielding high complexity. This section addresses the challenge of efficiently identifying optimal replication strategies. We first build a theoretical model to estimate parallel speedup from layer replication strategies, enabling rapid evaluation without physical cluster deployment. Building on this, we propose a scale-up algorithm that maximizes speedup while  minimizing communication overhead between discrete replicas. 

\textbf{Modeling of Speedup.} Considering that module replication introduces localized parallelism, we draw inspiration from Amdahl's Law \cite{Amdahl}. Its original formulation is $S(a,p) = \frac{1}{(1-a)+\frac{a}{p}}$, $a$ denotes the parallelizable fraction of the workload and $p$ denotes the degree of parallelism. However, for our scenario, Amdahl's Law requires three key modifications: 

\begin{itemize}[leftmargin=2em]
\item[(a)] Amdahl's Law assumes a uniform parallelism degree $p$ across all parallelizable layers. In contrast, replication allows each module $i$ to have a distinct parallelism degree $p_i$. We therefore represent scaling strategy using vector $P = [p_1, p_2, \ldots, p_n]$, where element $p_i$ denotes the replication factor or parallelism degree of module $i$, enabling unified computation of both parallel and sequential components. Additionally, $P_0 = [1, \ldots, 1] \in \mathbb{R}^n$ represents the initial fully sequential case representing the original configuration without parallelization.

\item[(b)] In Amdahl's Law, computational latency $w$ is canceled out in the fraction: $S(a,p) = \frac{w}{w \cdot (1-a) + w \cdot \frac{a}{p}}$. In our case, different strategies yield varying situations, so we explicitly model computation time through function $W(P)$.

\item[(c)] Amdahl's Law neglects communication overhead from parallelization, however replication introduce additional communication costs modeled by $T(P)$.
\end{itemize}

We introduce some essential notation: $bs_{ij}$ represents the batch size allocated to the $j$-th replica of the $i$-th layer.
$h$ represents the number of attention heads in the given instance, and $l$ represents the final sequence length for processing. $d$ represents the model dimension, which is the size of embeddings and hidden states throughout the model.  $B$ represents the network bandwidth of the cluster, where $B_{ij}$ represents the communication bandwidth between the $i$-th module in the original instance and its $j$-th replica. $C$ represents the computational capacity of the cluster, where $C_{ij}$ denotes the computing power allocated to the $j$-th replica of the $i$-th module in the original instance.

We first formalize $W(P)$, calculated as the accumulated computation load (per layer) divided by its hosting device's computing capacity:
\vspace{-0.3cm}
\begin{equation}
W(P) = \sum_{i=1}^{n} \max_{j=1}^{p_i} \frac{d^2 \times bs_{ij}
\times l
}{C_{ij}}.
\label{eq:comp_overhaed}
\end{equation}


Next, we formulate the communication overhead $T(P)$ during inference caused by replication operations. The communication volume corresponds to the batch size received by each replica, and the total number of communication events is therefore determined by the count of non-consecutive layer transitions based on \S \ref{sec:analysis_replication}, which we represent with the constant $\delta$ as:
\vspace{-0.4cm}
\begin{equation}
T(P) =  \delta\times \sum_{i=1}^n \sum_{j=1}^{p_i-1} \frac{d \times bs_{ij} 
\times l
}{B_{ij}}.
\label{eq:comm_overhead}
\end{equation}
It's important to note that, in order to simplify evaluation, $W(P)$ and $T(P)$ don't equal the actual concrete computation time and communication time but are positively correlated with them.

The speedup ratio $S(P)$ equals the fully sequential case $P_0$ divided by the sum of computation time and communication time under the current strategy:
\begin{equation}
S(P) = \frac{W(P_0)}{W(P) + T(P)}. 
\label{eq:speedup}
\end{equation}
If we adopt the strategy of evenly splitting batch sizes (which is the most common case) in clusters composed of homogeneous devices, we can remove the max function and some variable constraints, and obtain a concise form:
\begin{equation}
S_{homo}(P) = \frac{1}{\gamma+ \frac{1-\gamma}{n} \sum^{n}_{i=1} \frac{1}{p_i}},
\label{eq:speedup_homo}
\end{equation}
where $\gamma = \frac{\delta \cdot C}{d\cdot B}$ is a constant related to the cluster configuration. This formulation indicates that in a homogeneous cluster, the speedup exhibits a positive correlation with both the number of modules and the degree of parallelism, which aligns with the conclusions drawn in Section \ref{sec:analysis_replication}. 

By estimating the speedup of module replication strategies, this modeling enables \sys{}'s scale-up mechanism to evaluate strategy performance. 
The derived speedup formulation also remains consistent with our earlier findings. 

\textbf{Scale-Up Algorithm.} Algorithm \ref{alg:scale-up} demonstrates a scale-up algorithm, which aims to achieve maximum speedup while reducing communication frequency between discrete layers. The algorithm first computes the current speedup using Equation \ref{eq:speedup_homo} (or Equation \ref{eq:speedup} for heterogeneous clusters) through the coefficient $\gamma$ and the number of layers $n$. In this process, to adapt to programming languages, we use the L1 norm of the Hadamard quotient between the all-ones vector and $P$, denoted as $\|\mathbf{1} \oslash P\|_1$, to represent $\sum^{n}_{i=1} \frac{1}{p_i}$. Then, the algorithm iterates through all scalable devices which are filtered by $GetEligibleNodes(G)$ based on resource vacancy rate, and obtains the remaining resource capacity of the iterated device to determine the maximum number of replicas (max\_replicas) by dividing by the Replica Size $r$ (equivalent to the storage and computational requirements of a single layer). 

\begin{algorithm}[htbp]
\caption{Scale-Up Algorithm}
\label{alg:scale-up}
\small
\KwIn{LLM $model$, Cluster $G$, Current state $P$, Configuration Coefficient $\gamma$, Replica Size $r$, Layer Num $n$}
\KwOut{Optimal strategy $P^*$}
\BlankLine
$sp\_best \gets 1/({\gamma + (1-\gamma)/{n} * \|\mathbf{1} \oslash P\|_1})$\\
\For{$g_{dst} \in GetEligibleNodes(G)$}{
    $max\_replicas \gets g_{dst}.available / r$ \\
    $candidates \gets SortCandidatesByContinuity(P^*, g_{dst}, max\_replicas)$ \\
    \For{$layer\_id \in candidates$}{
        $P' \gets P^*.copy() $  \\
        $P'.addReplica(layer\_id,g_{dst}) $  \\
        $sp \gets 1/({\gamma + (1-\gamma)/{n} * \|\mathbf{1} \oslash P'\|_1})$ \\
        \If{$sp > sp\_best$}{
            $replicate(model,layer\_id,g_{dst})$ \\
            $sp\_best \gets sp$ \\
            $P^* \gets P'$
        }
    }
}
\Return{$P^*$}
\end{algorithm}

Next, to minimize communication frequency between discrete layers, the \textit{SortCandidatesByContinuity()} function collects layers that can be replicated on the current device, uses continuity as the sorting criterion to prioritize layer IDs for replication, and returns the top max\_replicas candidates. Specifically, continuity means the longest continuous sequence of layer indices receives the highest priority, and when all replicas form continuous sequences, priority aligns with the layer indices. For each selected candidate layer, the algorithm simulates the speedup improvement after adding a replica. Only if the new configuration provides a higher speedup does it actually execute the replica addition operation.

The proposed scale-up algorithm efficiently navigates the search space of layer replication strategies while maintaining two key properties: (a) it guarantees monotonic improvement in speedup through greedy selection of optimal local configurations, and (b) it preserves communication efficiency by preferentially maintaining continuous layer sequences.

Building on this foundation, we develop a scale-up algorithm that quickly explores replication strategies by prioritizing continuous layer placement and incremental speedup improvements. This approach effectively balances search efficiency with solution quality while maintaining compatibility with \sys{}'s existing resource management framework. The subsequent evaluation will demonstrate how this mechanism operates in practical deployment scenarios.


\vspace{-0.3cm}
\subsection{Scale-Down: Robustness via Module Reduction}

When workload intensifies beyond capacity, causing the SLO violation rate to exceed predetermined thresholds or even triggering OOM failures, \sys{} necessitates scale-down operations for LLM serving. To enhance robustness against SLO violations and OOM failures, we propose Module Reduction, a three-phase intervention approach that executes sequentially until the violations are sufficiently reduced:
\begin{itemize}[leftmargin=2em]
\item[(a)] \textbf{Module Migration}: Initially, \sys{} attempts migration to fine-grained optimize instance deployment in order to mitigate computational load and memory pressure across devices.
\item[(b)] \textbf{Replica Eviction}: In scenarios where migration cannot resolve the issue, the system identifies and sequentially removes layer replicas co-located with the affected model, prioritizing those with minimal impact on overall serving performance.
\item[(c)] \textbf{Performance Reduction}: When prior interventions prove insufficient, \sys{} incrementally reduces the model's batch size and try offloading to solve issues. 
\end{itemize}

\begin{algorithm}[htbp]
\caption{Scale-down Algorithm}
\label{alg:scale_down}
\small
\KwIn{LLM $model$, Cluster $G$, Current state $P$, SLO threshold $\theta$, Batch size $bs$, Adjustment step $\Delta bs$}
\KwOut{Updated state $P'$, Updated batch size $bs'$}

$g_{src} \gets model.device()$\\
$P' \gets P$\\
// Phase 1: Module Migration\\
$modules \gets FilterModules(model, g_{src})$\\
\For{$m \in modules$}{
    $g_{dst} \gets FindOptimalDestination(G, m)$\\
    $migrate(m, g_{src}, g_{dst})$\\
    \If{$\neg IsViolating(g_{src}, \theta, P')$}{
        \textbf{return} $P', bs$
    }
}

// Phase 2: Replica Eviction\\
$evictees \gets SortEvicteesBy(P, g)$ \\
\For{$layer\_id \in evictees$}{
    $P' \gets P'.evictReplica(model, layer\_id, g)$ \\
    \If{$\neg IsViolating(g, \theta, P')$}{
        \textbf{return} $P', bs$
    }
}

// Phase 3: Performance Reduction\\
\While{$IsViolating(G, \theta, P)$ \textbf{and} $bs >= 1$}{
    $bs' \gets bs - \Delta bs$\\
    $ReduceBatchSize(model, bs')$\\
    $executeOffloading(model)$
    \If{$\neg IsViolating(G, \theta, P')$}{
        \textbf{break}
    }
    $bs \gets bs'$
}

\textbf{return} $P, bs$
\end{algorithm}

We implement Module Reduction in Algorithm \ref{alg:scale_down}. The three actions are executed sequentially, with progressively increasing costs. In the first phase, Module Migration, the algorithm filters modules that can alleviate computational load or memory pressure on the source device through \textit{FilterModules()}. This filtering process determines the number of candidates based on the analysis in \S \ref{sec:analysis_migration}, rather than returning full model. For each candidate module, it determines an optimal destination device with sufficient resources and performs migration. If SLO violations are resolved after migration, the algorithm terminates and returns the updated state.

When violations persist, the algorithm advances to the second phase, Replica Eviction. During this phase, it makes best effort to evict replicas, based on the priority outlined in \S \ref{alg:scale-up} as the order of eviction. This intermediate phase accepts a moderate performance trade-off to restore system stability.

When all scaling operations fail, we can only ensure the SLO attainment by reducing the original performance. The algorithm gradually reduces the batch size through \textit{ReduceBatchSize()} by an appropriate step size $\Delta bs$ (e.g., 5) and attempts to alleviate the problem by offloading parts of the model such as parameters and KV Cache.

The algorithm's graduated intervention approach ensures that remediation strategies with lower performance impacts are exhausted before more costly measures are implemented, thereby maintaining optimal service quality, even in challenging operational conditions.


\section{CoCoServe System and Implementation}
\label{sec:sys}
\begin{figure}[htbp]
    \centering
    \includegraphics[width=0.9\linewidth]{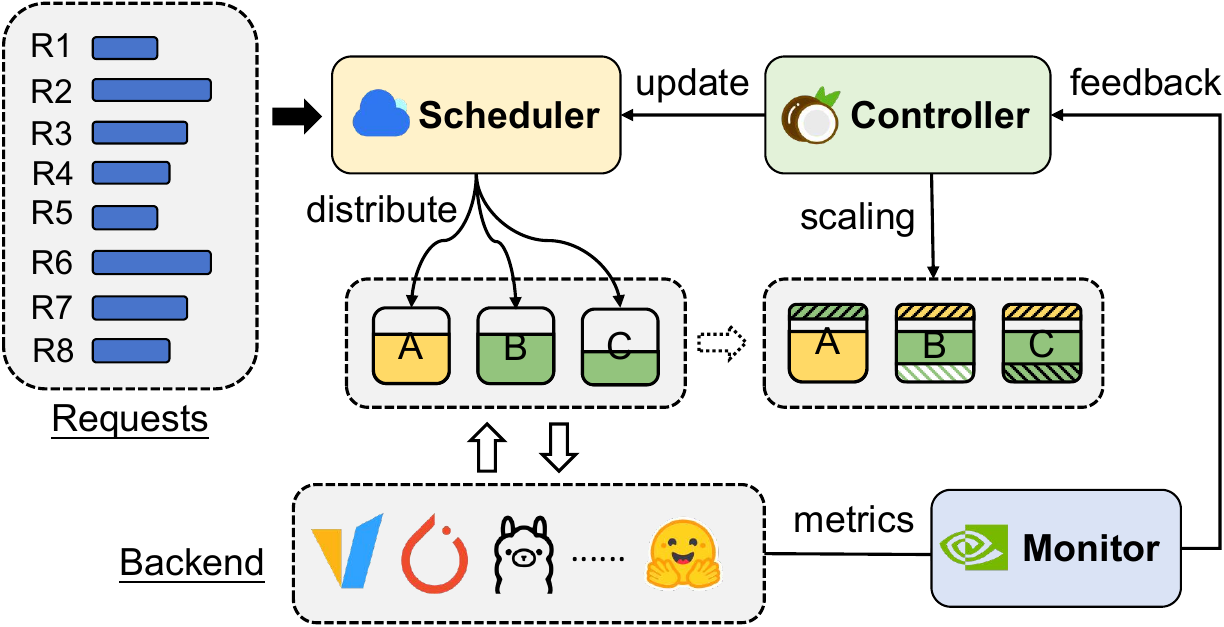}
    \caption{\sys{}'s component interaction sequence.}
    \label{fig:sys}
\end{figure}

\sys{} is an efficient serving system designed for the management and scaling of LLM services with the aforementioned techniques in Sections \ref{sec:operation} and \ref{sec:mechanism}. As illustrated in Fig. \ref{fig:sys}, the system consists of several key components. The \textit{Scheduler} receives incoming requests and distributes them across different model instances. A \textit{Controller} component works in conjunction with the \textit{Scheduler}, implementing scaling decisions based on real-time feedback. The system incorporates a \textit{Monitor} component that collects various performance and utilization metrics from the backend LLM engines~\cite{sosp23-vllm,wolf-etal-2020-transformers,xFormers2022,ollama}, which support multiple serving systems. These metrics are then fed back to the \textit{Controller}, forming a closed control loop. When scaling is required, the \textit{Controller} triggers appropriate actions on the modules, as shown by the hatched blocks in the rightmost deployment group. This architecture enables dynamic resource allocation and efficient request handling while maintaining system stability through continuous monitoring and feedback-driven control.

\textbf{Metrics Monitor.} The monitor collects metrics including GPU utilization, memory utilization, tokens per second, end-to-end latency, etc., and feeds these metrics back to the \textit{Controller}. For utilization metrics, the monitor gathers data via the NVML (NVIDIA Management Library)~\cite{NVML} interface that enables querying real-time telemetry. For performance metrics, the monitor directly adopts the data provided by the backend LLM engine; if the backend does not support this, the monitor will actively add system timers, such as \textit{time.perf\_counter()} and \textit{event.record()}, which is consistent with the approach used by mainstream engines.

\textbf{Auto-Scaling Controller.} The \textit{Controller} serves as the core component for auto-scaling operations. It periodically collects system performance metrics from the \textit{Monitor} module and evaluates potential scaling strategies according to the workflow defined in Section \ref{sec:mechanism}. The controller employs two key thresholds to determine scaling actions: it triggers scale-up operations when the resource vacancy rate exceeds $T_{up}$, and initiates scale-down operations when the SLO violation rate surpasses $T_{down}$. Upon finalizing its decision, the \textit{Controller} executes the predetermined operations on target modules. Following each scaling operation, it updates the \textit{Scheduler} with the latest instance information. To prevent GPU memory and computational overload, the \textit{Controller} implements threshold-based safeguards. When initiating scaling operations, it performs a  evaluations of all available GPUs' current states to select the most suitable target device based on resource availability and workload distribution.

\textbf{Request Scheduler.} The \textit{Scheduler} is responsible for efficiently distributing incoming inference requests from users to backend instances. It allocates requests based on the current workload distribution across instances and the updated instance performance provided by the \textit{Controller}. Furthermore, the \textit{Scheduler}'s implementation is inherited from the backend LLM engine, supporting features such as continuous batching~\cite{osdi22-orca}, streaming scheduling and others~\cite{icsoc24-uellm}.

\begin{figure*}[hpt]
     \centering
    \begin{subfigure}[b]{0.247\textwidth}
        \centering
        \includegraphics[width=\linewidth]{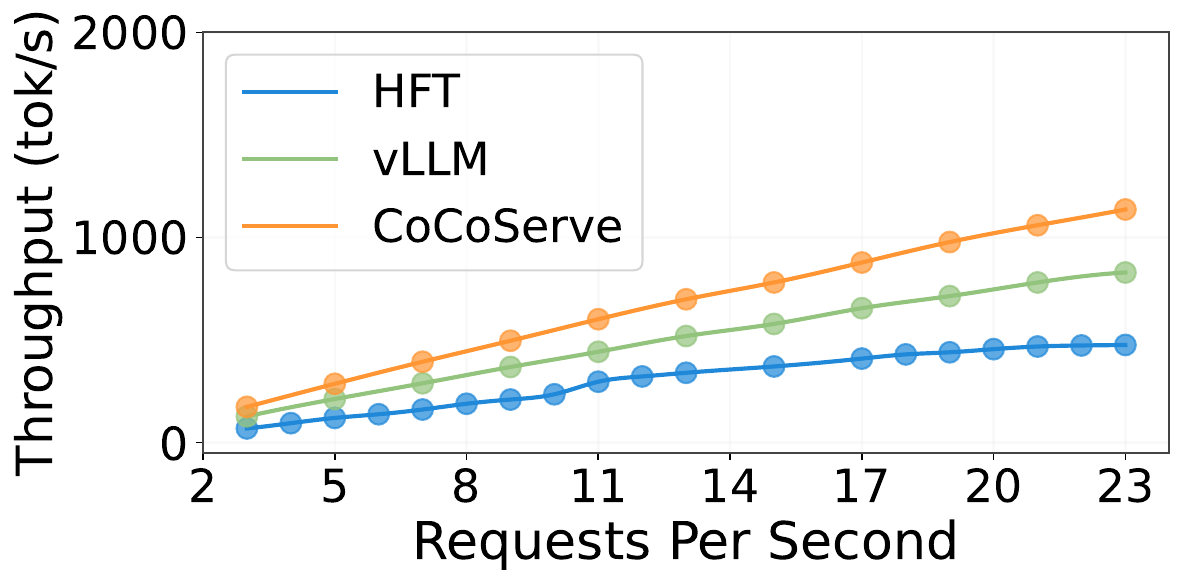}
        \caption{13B Throughput (Low Load)}
        \label{fig:13b_tp_low}
    \end{subfigure}
    \hfill
    \begin{subfigure}[b]{0.24\textwidth}
        \centering
        \includegraphics[width=\linewidth]{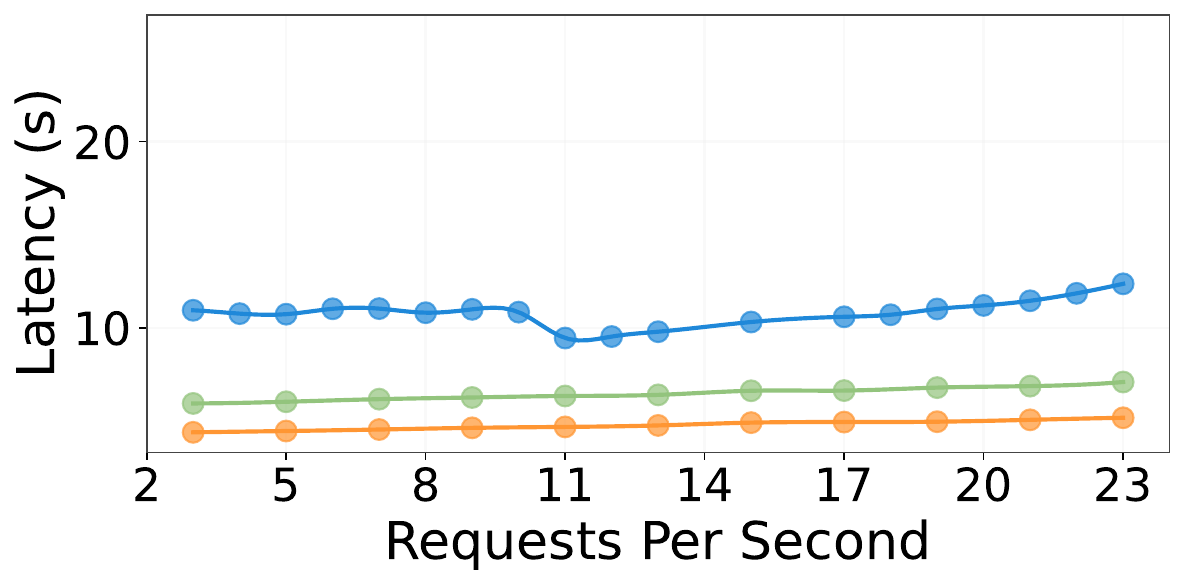}
        \caption{13B Latency (Low Load)}
        \label{fig:13b_lat_low}
    \end{subfigure}
    \hfill
    \begin{subfigure}[b]{0.24\textwidth}
        \centering
        \includegraphics[width=\linewidth]{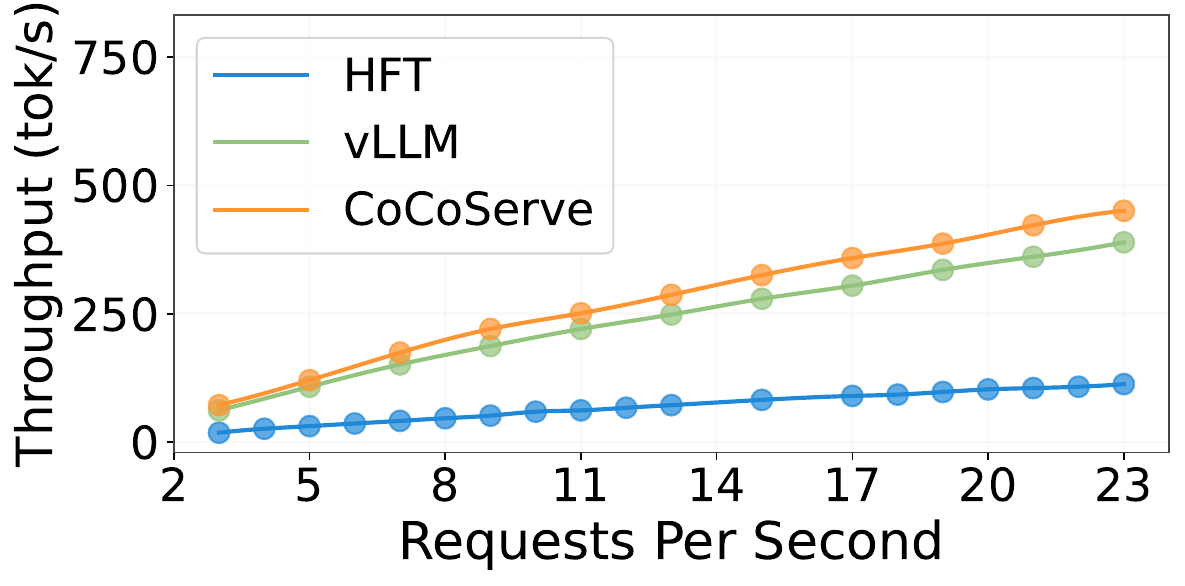}
        \caption{70B Throughput (Low Load)}
        \label{fig:70b_tp_low}
    \end{subfigure}
    \hfill
    \begin{subfigure}[b]{0.24\textwidth}
        \centering
        \includegraphics[width=\linewidth]{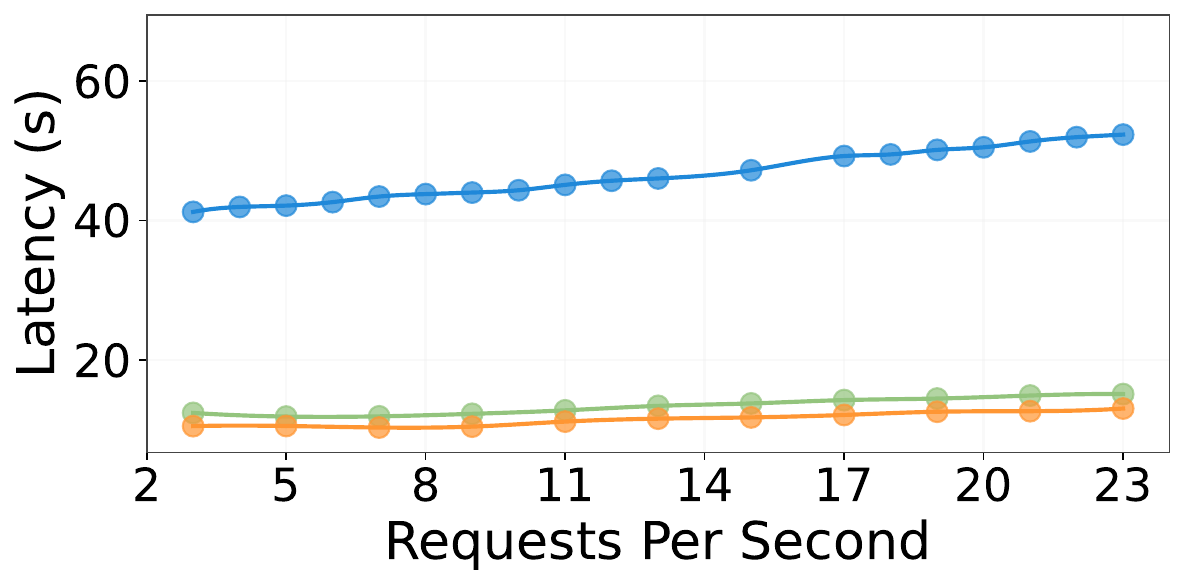}
        \caption{70B Latency (Low Load)}
        \label{fig:70b_lat_low}
    \end{subfigure}
    
    \begin{subfigure}[b]{0.247\textwidth}
        \centering
        \includegraphics[width=\linewidth]{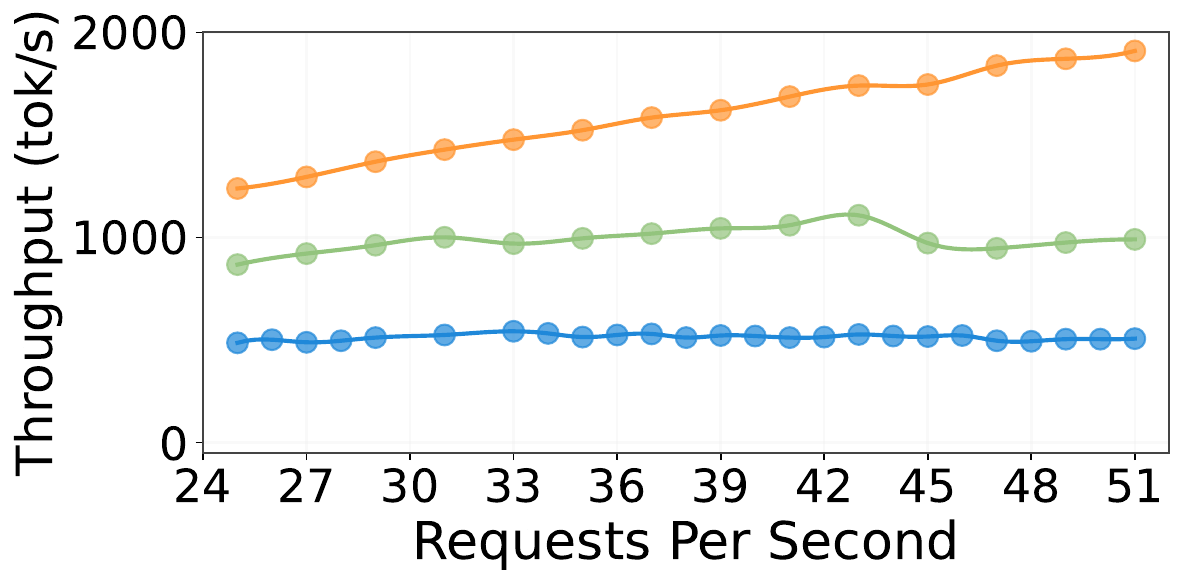}
        \caption{13B Throughput (High Load)}
        \label{fig:13b_tp_high}
    \end{subfigure}
    \hfill
    \begin{subfigure}[b]{0.24\textwidth}
        \centering
        \includegraphics[width=\linewidth]{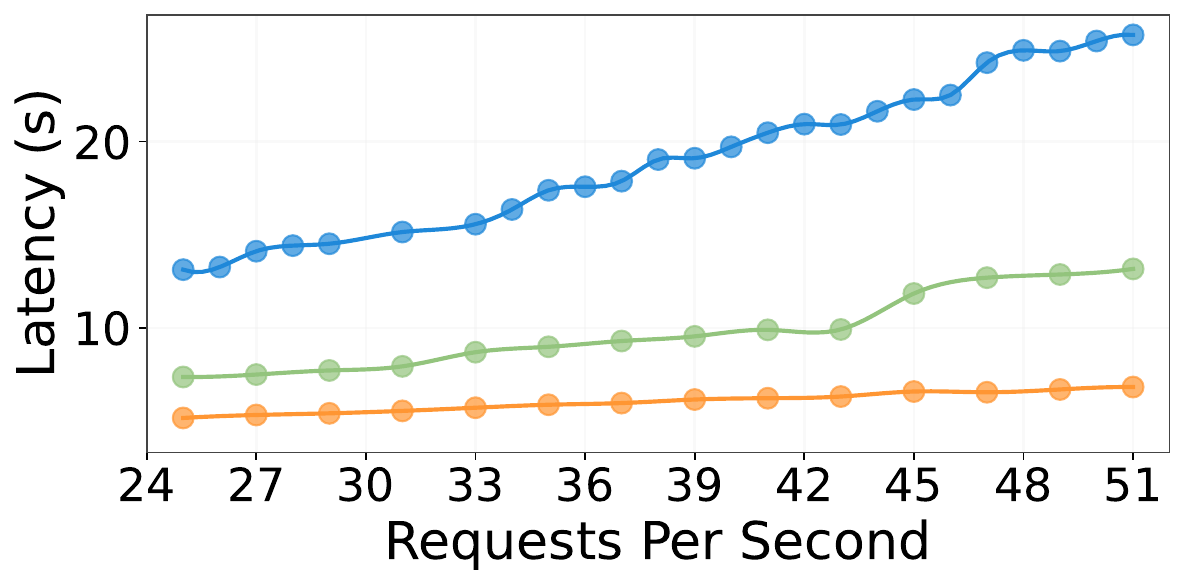}
        \caption{13B Latency (High Load)}
        \label{fig:13b_lat_high}
    \end{subfigure}
    \hfill
    \begin{subfigure}[b]{0.24\textwidth}
        \centering
        \includegraphics[width=\linewidth]{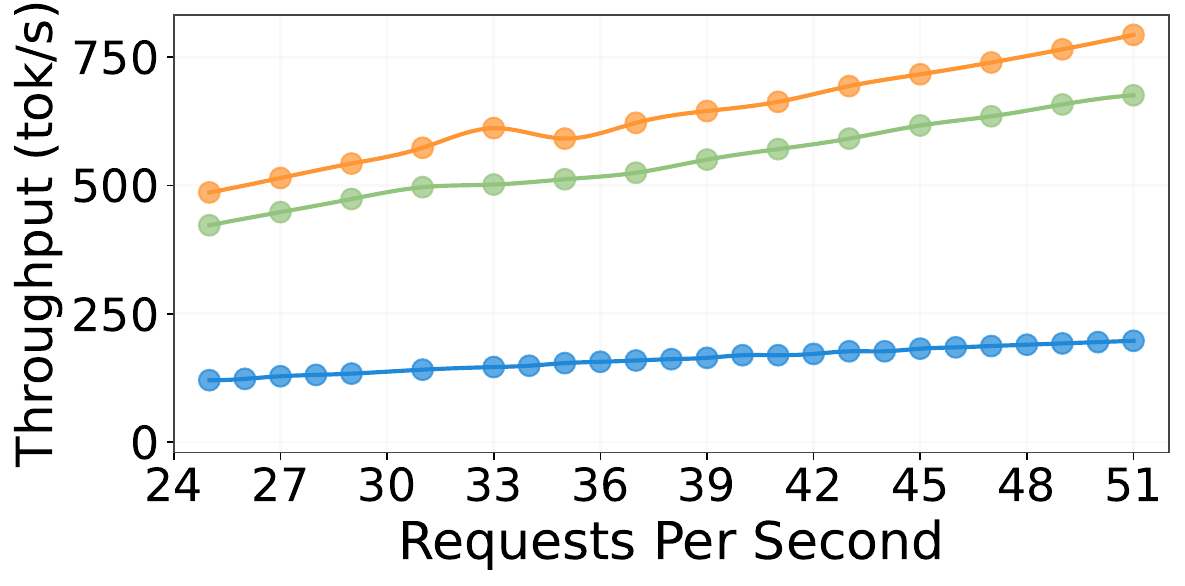}
        \caption{70B Throughput (High Load)}
        \label{fig:70b_tp_high}
    \end{subfigure}
    \hfill
    \begin{subfigure}[b]{0.24\textwidth}
        \centering
        \includegraphics[width=\linewidth]{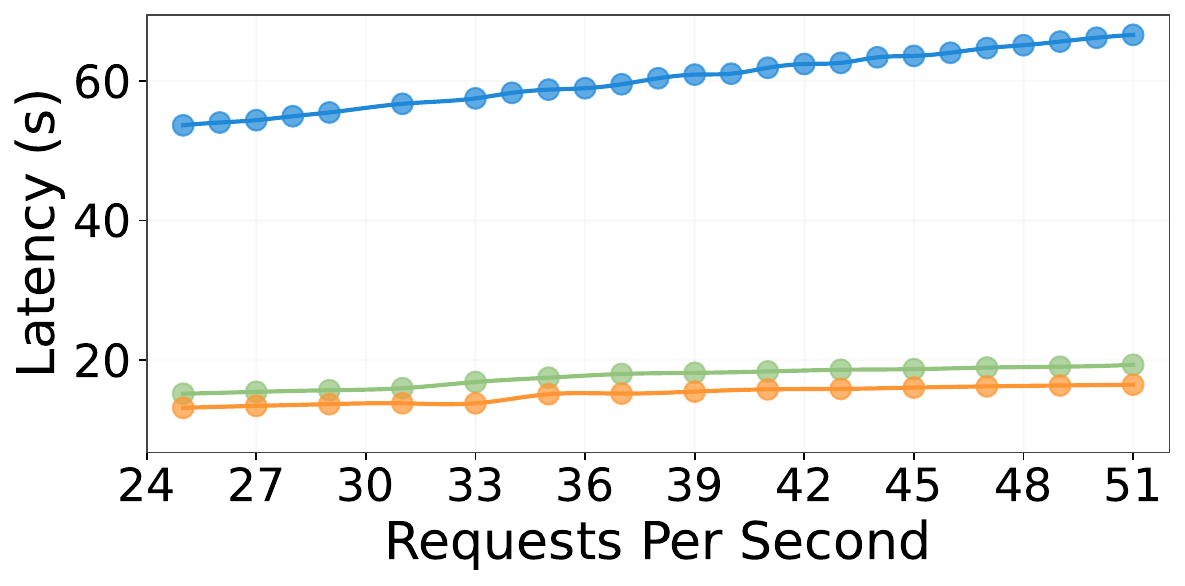}
        \caption{70B Latency (High Load)}
        \label{fig:70b_lat_high}
    \end{subfigure}
    
    \caption{Throughput and latency comparison between CoCoServe, \hft{}, and vLLM for a single LLaMA-13B and LLaMA-70B instance deployed on an A100 GPU under different load conditions.}
    \label{fig:e1}
\end{figure*}

\section{Performance Evaluations}
\label{sec:expr}
In this section, we evaluate \sys{} from the perspectives including resource consumption, SLO attainment, OOM and operational costs under different workloads compared with state-of-the-art systems.

\vspace{-0.1cm}
\subsection{Experiment Setup}
Our experiments were conducted on a server equipped with four NVIDIA A100-40GB PCIe GPUs. We evaluated our system, CoCoServe, on both LLaMA2-13B and LLaMA2-70B models, using \HFT{} 4.51 and vLLM 0.8.5 as our comparison targets, which represent the state-of-the-art in LLM serving systems. All experiments utilized BF16 precision to achieve an optimal balance between computational efficiency and numerical accuracy. We configured the maximum sequence length for token generation at 256 tokens and employed the Alpaca dataset~\cite{23-alpaca}. To evaluate the performance of CoCoServe under different workload levels, we conducted testing across two distinct workload environments: low (3-30 RPS) and high (31-50 RPS). For each request rate, we repeated the tests five times to remove randomness.

\vspace{-0.1cm}
\subsection{Single LLM Instance Case}
\label{sec:expr1}

Fig. \ref{fig:e1} shows performance comparisons across different load conditions for one deployed instance each of LLaMA-13B and LLaMA-70B models, measuring throughput and latency. 

Under low workload conditions, CoCoServe shows remarkable efficiency gains with both tested models. The LLaMA-13B model achieves, on average, 56.89\% latency reduction and 2.13$\times$ throughput compared to \hft{} systems. When compared to vLLM, it still maintains 26.87\% lower latency and 36.89\% higher throughput on average. The larger LLaMA-70B model shows even more impressive results under low workloads, with average latency reductions of 75.22\% and throughput improvements of 4$\times$ compared to \hft{} systems, while maintaining a 13.65\% latency advantage and 15.56\% throughput improvement over vLLM on average.

Under high workload conditions, \hft{} systems experience dramatic performance degradation. With the 70B model, CoCoServe demonstrates substantial advantages, reducing latency by 74.85\% and improving throughput by 3.9$\times$ compared to \hft{} systems. While both CoCoServe and vLLM maintain relatively stable performance under increased workloads, CoCoServe consistently outperforms vLLM with 14.67\% lower latency and 17.14\% higher throughput on average, highlighting its superior efficiency in handling demanding computational scenarios.

\begin{figure}[htbp]
    \centering
    \includegraphics[width=0.7\linewidth]{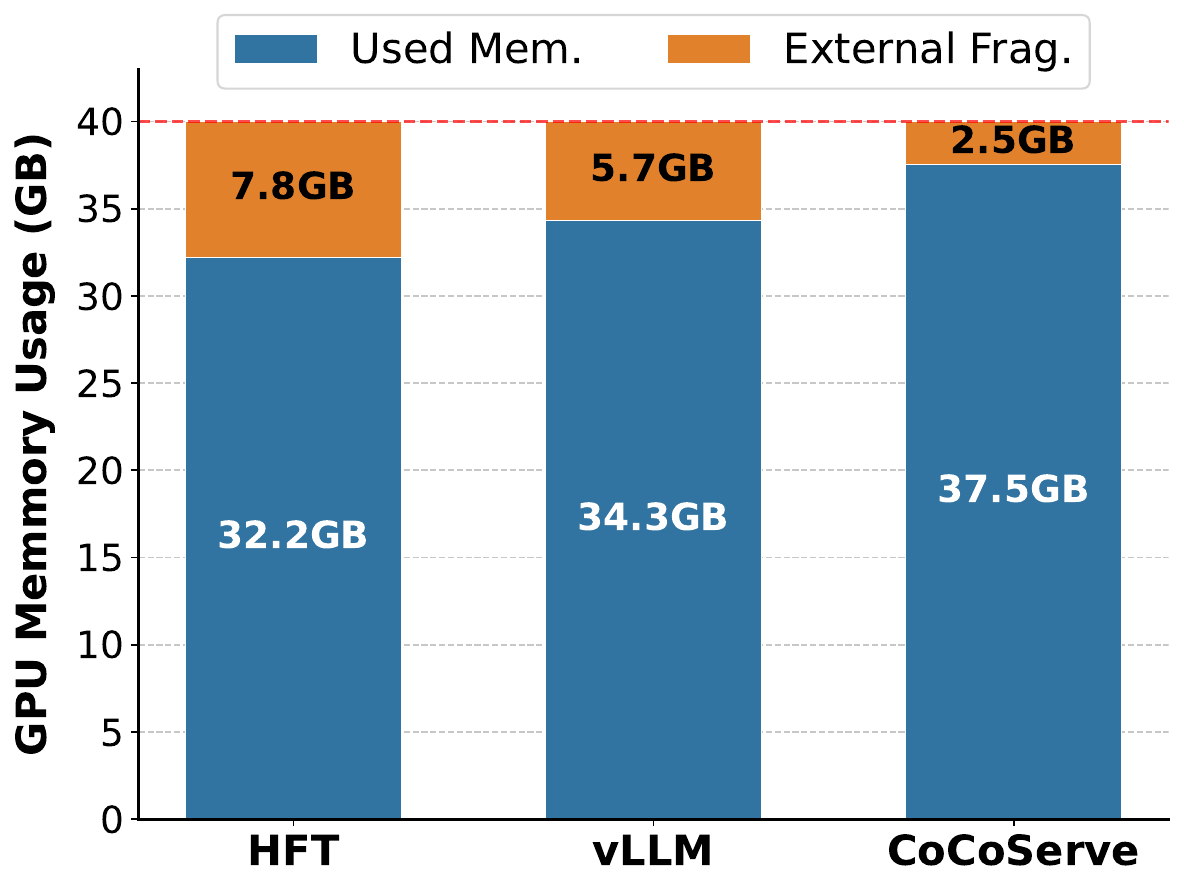}
    \caption{Memory utilization comparison.}
    \label{fig:e1_utilize}
\end{figure}

In terms of GPU memory efficiency, as shown in Fig. \ref{fig:e1_utilize}, CoCoServe demonstrates significant memory optimization, achieving 5.3GB (vs. HFT) and 3.2GB (vs. vLLM) reductions in wasted memory. The system reduces memory fragmentation by 3.12$\times$ compared to HFT and 2.28$\times$ relative to vLLM. This enables CoCoServe to effectively utilize 37.5GB of memory for model serving - surpassing HFT's capacity by 5.3GB and vLLM's by 3.2GB under identical hardware constraints.

These results demonstrate that CoCoServe consistently outperforms both \hft{} and vLLM systems across all tested scenarios. In summary, under various RPS test scenarios, CoCoServe demonstrates significant performance advantages. 
The combined benefits of reduced latency, increased throughput, and optimized memory utilization establish CoCoServe as a highly efficient serving system for large language models, particularly when handling complex workloads or operating with limited computational resources.
\vspace{-0.35cm}
\subsection{Multiple LLM Instances Case}
\label{sec:expr2}
To evaluate CoCoServe's performance in multiple LLM instance scenarios, we deployed various configurations on our 4$\times$A100 GPUs: 2 instances using CoCoServe, 2 \hft{} instances, and 4 \hft{} instances. 
Each instance ran a 13B model with a maximum generation length of 256 tokens, enabling us to directly compare performance differences between these deployment configurations. Given that our previous experiment had already demonstrated CoCoServe's and \hft{}'s superior performance over vLLM in single-instance scenarios, we focused our multi-instance comparison solely on \hft{} instances as the baseline to evaluate the scalability advantages of our approach.

\begin{figure}[htbp]
    \centering
    \begin{subfigure}[b]{0.48\linewidth}
        \centering
        \includegraphics[width=\linewidth]{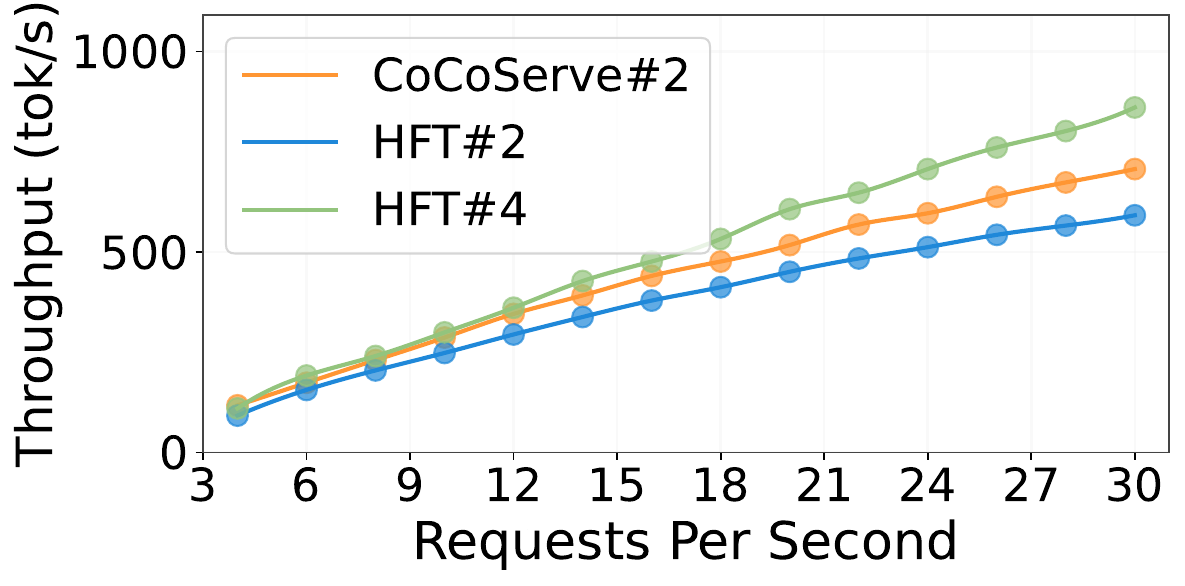}
        \caption{Throughput (Low Load)}
        \label{fig:e2a}
    \end{subfigure}
    \hfill
    \begin{subfigure}[b]{0.48\linewidth}
        \centering
        \includegraphics[width=\linewidth]{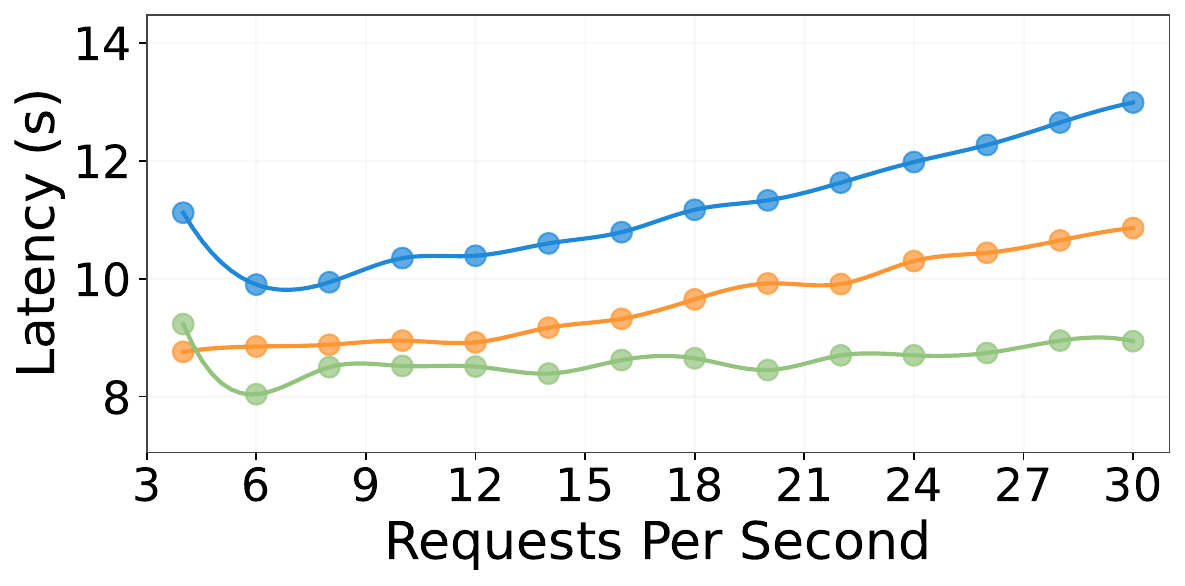}
        \caption{Latency (Low Load)}
        \label{fig:e2b}
    \end{subfigure}
    \hfill

    \begin{subfigure}[b]{0.48\linewidth}
        \centering
        \includegraphics[width=\linewidth]{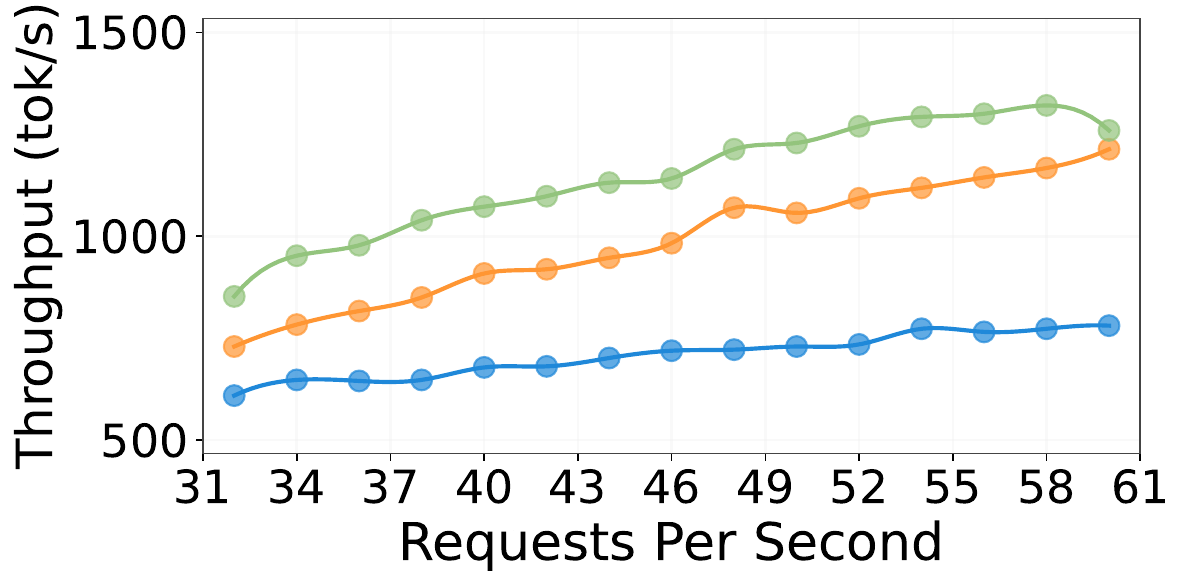}
        \caption{Throughput (High Load)}
        \label{fig:e2c}
    \end{subfigure}
    \hfill
    \begin{subfigure}[b]{0.48\linewidth}
        \centering
        \includegraphics[width=\linewidth]{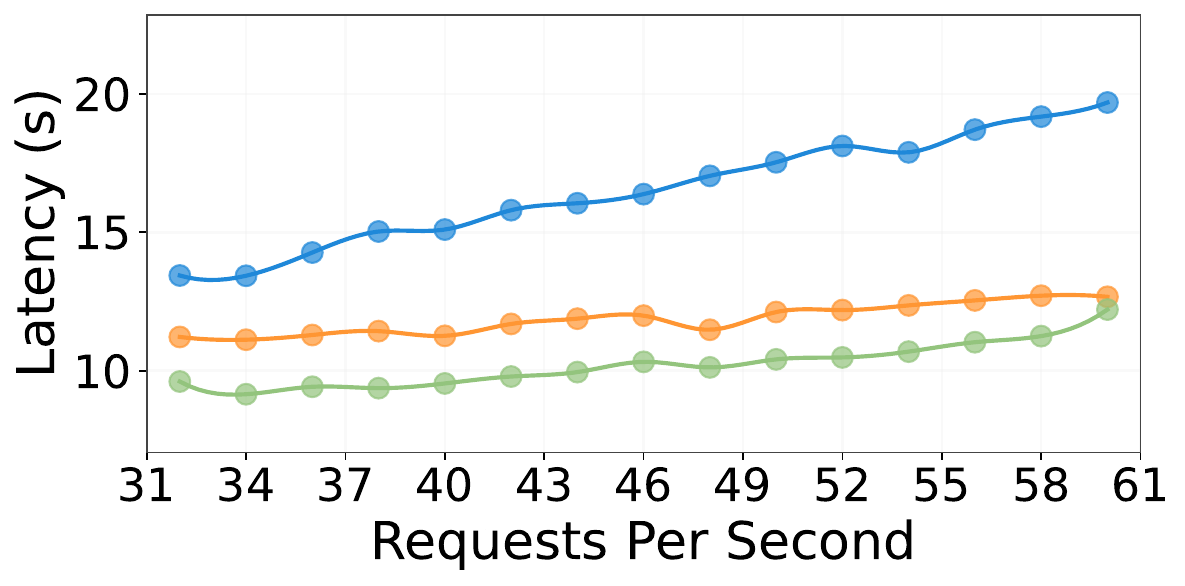}
        \caption{Latency (High Load)}
        \label{fig:e2d}
    \end{subfigure}

\caption{Throughput and latency comparison between CoCoServe (2 LLM instances) and HFT (2 and 4 LLM instances) for LLaMA-13B and LLaMA-70B models deployed on an A100 GPU under different load conditions.}
\label{fig:e2}
\end{figure}

As shown in Fig. \ref{fig:e2a} and \ref{fig:e2b}, CoCoServe demonstrates significant efficiency advantages in low workload scenarios. When compared to the 2-instance \hft{} configuration, CoCoServe achieves consistent latency reductions averaging 14.23\% and throughput improvements of approximately 16.92\% across various request rates. The performance gap widens as the request rate increases, with latency reductions reaching higher values at increased request rates. While the 4-instance \hft{} configuration does show better raw performance than CoCoServe's 2-instance deployment, the difference is relatively modest (approximately 11.28\% better latency and 9.59\% better throughput on average) despite using twice the number of instances. This highlights CoCoServe's exceptional resource efficiency, delivering comparable performance with half the deployment resources.

The advantages of CoCoServe become even more pronounced under high workload conditions, as illustrated in Figs. \ref{fig:e2c} and \ref{fig:e2d}. When compared to the 2-instance \hft{}, CoCoServe delivers dramatic performance improvements with latency reductions averaging 27.48\% and throughput gains averaging 38.78\%. 

While the 4-instance \hft{} configuration maintains an edge over CoCoServe's 2-instance deployment under high workloads (16.36\% better latency and 13.94\% better throughput on average), the performance difference narrows significantly as request rates increase. At the highest tested loads, the difference becomes smaller, despite CoCoServe using only half the resources compared with \hft{}.

Moreover, the resource utilization measurements reveal important efficiency differences between the configurations. The 2-instance \hft{} configuration requires 58,786.68 MiB of GPU memory, while the 4-instance \hft{} configuration consumes 119,573.34 MiB. In comparison, CoCoServe's 2-instance deployment utilizes 64,015.44 MiB, which represents only a 9\% increase over the 2-instance \hft{} while achieving substantially better performance. Most notably, CoCoServe's memory footprint is only 53.54\% of the 4-instance \hft{} configuration, reducing cost by over 46\%, while delivering approximately 90\% of its performance. This demonstrates CoCoServe's remarkable efficiency in resource utilization, providing near-equivalent performance at approximately half the computational cost.

The multi-instance evaluation clearly demonstrates CoCoServe's advantages in serving LLM efficiently. When compared to \hft{} systems with the same number of instances, CoCoServe delivers consistent performance improvements across all workload conditions. CoCoServe's remarkable efficiency in resource utilization, providing near-equivalent performance at approximately half the computational cost.

\subsection{Robustness Validation}
\label{sec:expr3}

\begin{figure}[htbp]
    \centering
    \begin{subfigure}[h]{0.45\linewidth}
        \centering
        \includegraphics[width=\linewidth]{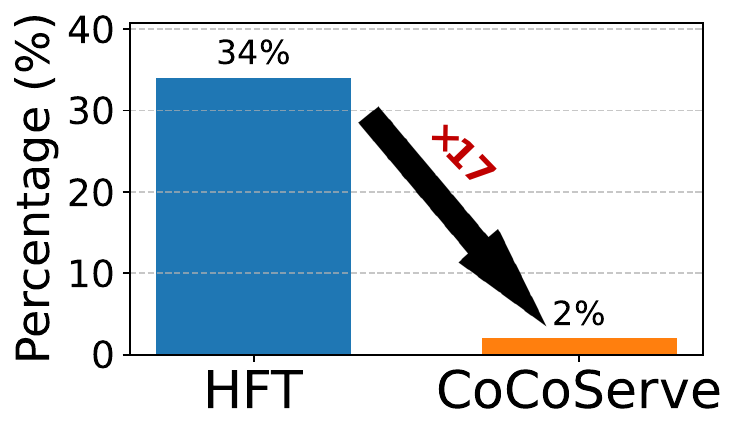}
        \caption{OOM occurrence rate comparison between HFT and CoCoServe systems}
        \label{fig:e3a}
    \end{subfigure}
    \hfill
    \begin{subfigure}[h]{0.45\linewidth}
        \centering
        \includegraphics[width=\linewidth]{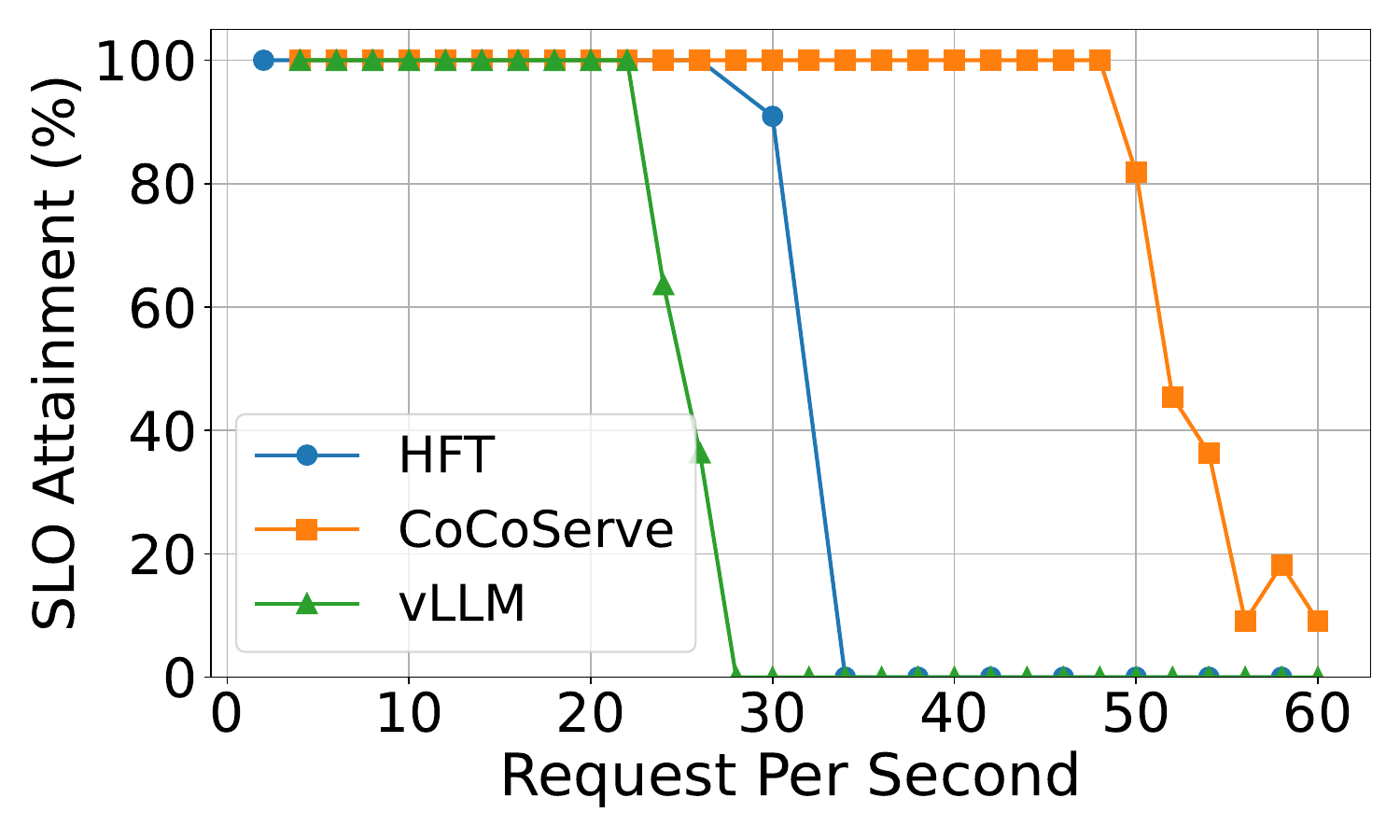}
        \caption{SLO attainment comparison among HFT, CoCoServe, and vLLM systems}
        \label{fig:e3b}
    \end{subfigure}
    \hfill
\label{fig:e3}
\caption{Performance comparison between serving systems: (a) shows CoCoServe reduces OOM occurrences by 17× compared to HFT, while (b) demonstrates superior SLO attainment of CoCoServe compared to HFT and vLLM.}
\end{figure}

Our experiments also revealed significant differences in memory usage between the systems. As shown in Fig. \ref{fig:e3a}, the \hft{} system exhibited a concerning 34\% OOM error rate during high-load scenarios, particularly when request rates exceeded 50 requests per second. In contrast, CoCoServe demonstrated remarkable stability with only a 2\% OOM occurrence rate under identical conditions, which represents a 17$\times$ improvement in memory stability.

Fig. \ref{fig:e3b} illustrates the SLO attainment rates across different request loads for \hft{}, CoCoServe, and vLLM systems. The results clearly demonstrate CoCoServe's superior robustness under increasing workloads. While the \hft{} system's SLO attainment begins to deteriorate at approximately 25 requests per second and completely fails beyond 30 requests per second, CoCoServe maintains perfect SLO attainment until approximately 50 requests per second.


These results validate CoCoServe's enhanced robustness, providing strong evidence of its capability to handle unexpected traffic surges and maintain service quality under challenging conditions. The dramatic reduction in OOM failures combined with extended SLO attainment makes CoCoServe particularly valuable for mission-critical applications where service interruptions must be minimized.

\subsection{Scalability Analysis with Scaling Cost}
\label{sec:expr4}

\begin{table}[htbp]
\centering
\caption{Replication and Migration Cost Analysis}
\setlength{\abovecaptionskip}{0cm} 
\setlength{\belowcaptionskip}{-1cm}
\small 
\setlength{\tabcolsep}{4pt} 
\begin{tabular}{c|cc|cc}
\hline
\multirow{2}{*}{\textbf{No. of Layers}} & \multicolumn{2}{c|}{\textbf{Replication}} & \multicolumn{2}{c}{\textbf{Migration}} \\
\cline{2-5}
 & \textbf{Time} & \textbf{Memory} & \textbf{Time} & \textbf{Memory} \\
\hline
1  & 0.2987 s & 1107 MB & 0.2492 s & 1107 MB \\
10 & 0.3581 s & 6579 MB & 0.3181 s & 6579 MB \\
20 & 0.3826 s & 12659 MB & 0.3426 s & 12659 MB \\
30 & 0.4947 s & 18739 MB & 0.3947 s & 18739 MB \\
40 & 0.8938 s & 24819 MB & 0.8138 s & 24819 MB \\
\hline
\end{tabular}
\label{tb:cost}

\end{table}

To validate \sys{}'s scalability in terms of scaling costs, Table \ref{tb:cost} presents a comprehensive analysis of the time and memory costs associated with model replication and migration operations. As the number of manipulated model layers increases from 1 to 40, we observe a gradual increase in both time and memory requirements. For replication operations, the time cost increases from 0.2987 seconds for a single layer to 0.8938 seconds for 40 layers, representing a modest 3$\times$ increase despite a 40$\times$ scaling in model count. Similarly, migration operations show slightly better efficiency, with time costs ranging from 0.2492 seconds to 0.8138 seconds across the same scaling range.

Memory consumption scales linearly with the number of layers, from 1107 MB for a single layer to 24819 MB for 40 layers. This linear relationship indicates predictable resource requirements as the system scales. Notably, migration operations maintain the same memory usage as replication operations, suggesting efficient memory management during layer transfers.

After executing model scaling operations, communication coordination between replicas is necessary. Our experiments indicate that the time cost for this communication is 39.1 ms, while the additional memory consumption is negligible (with measurement values extremely small). This suggests that the inter-replica communication protocol is highly efficient and does not constitute a bottleneck for system scaling.

Overall, these results demonstrate that CoCoServe maintains remarkably low scaling costs even as the system expands to accommodate larger workloads. The sub-second time costs for both replication and migration operations, combined with the minimal communication overhead, enable rapid adaptation to changing demand patterns without significant performance penalties. This efficient scaling behavior is particularly valuable in dynamic serving environments where resource utilization must be continuously optimized.

\section{Related Work}
\label{se:related}
\vspace{0.2cm}
\textbf{Parallelism.} The replication operation in \sys{} introduces localized parallelism, a specialized variant of data parallelism. Mainstream model parallelism\cite{eurosys25-MEPipe,nips18-gpipe,eurosys25-mist} techniques for LLM architectures encompass Tensor Parallelism (TP), Pipeline Parallelism (PP), Data Parallelism (DP), and Sequence Parallelism (SP)~\cite{Wu2024LoongServeES}. These methods achieve parallelism by partitioning computations at distinct granularities—within layers, across layers, across data batches, and within sequences, respectively. Given the orthogonal nature of these parallelization approaches, recent research has explored the trade-offs between them to enable automatic configuration optimization. Megatron-LM~\cite{arxiv19-megatron} implements an effective hybrid of PP, TP, and DP, thereby enhancing scaling efficiency across diverse computational resources. Alpa~\cite{osdi22-alpa} further investigates the hierarchical configuration space of PP and TP building upon Megatron-LM and, leveraging this insight, achieves automatic parallelization reconfiguration. Notably, in contemporary LLM serving systems, the effect of data parallelism closely resembles load balancing across multiple instances, and its acceleration efficiency frequently surpasses other parallelism strategies due to reduced communication overhead, albeit at the cost of additional resource replication. In contrast to these methods, the module-level replication in \sys{} achieves partial data parallelism through layer-level replication rather than instance-level replication, rendering it highly cost-efficient in resource-constrained environments by precisely controlling resource allocation and effectively utilizing otherwise idle computational resources.

\noindent
\textbf{Offloading.} The migration operation in \sys{} shares conceptual similarities with offloading methods~\cite{eurosys25-hcache,arxiv-hostCaching} but differs fundamentally in its approach and underlying objectives. In response to the substantial memory requirements of LLMs, recent research has developed various efficient offloading strategies to mitigate resource constraints. DeepSpeed~\cite{kdd20-deepspeed} proposes a sophisticated multi-tier offloading scheme capable of progressively offloading parameters and intermediate variables to main memory or even secondary storage devices. InfiniGen~\cite{osdi24-infinigen} significantly reduces offloading overhead by selectively prefetching only essential KV cache entries rather than retrieving the entire cache. In contrast, module-level migration in \sys{} operates at the granularity of functional modules, which can range from entire decoder layers to finer-grained components such as projection matrices and KV caches. While both migration and traditional offloading techniques effectively alleviate computational and storage pressure on the source device, the primary objective of migration in \sys{} is to achieve a more optimal deployment configuration under the current workload conditions—typically manifested as improved load balancing and a more equitable distribution of compute and memory resources across the available infrastructure.

\section{Discussions}
\vspace{0.2cm}
\textbf{Compatibility and Scalability}. Module-level Migration and Replication are orthogonal to mainstream parallel methods, as they only change the location of modules and the sharding of data. Furthermore, the auto-scaling mechanism proposed by CoCoServe is also compatible with mainstream inference frameworks like vLLM and xFormers. However, since replication dynamically alters the data flow, it conflicts with the graph mode functionality. Additionally, CoCoServe can support scenarios with heterogeneous devices, and module-level scaling can more accurately utilize the environments provided by each device, achieving an ideal balance.

\noindent \textbf{Interference and Accuracy}. To test the interference of scaling operations on other tasks in multi-instance scenarios, we conducted tests on CoCoServe. Experimental data shows that during the execution of dynamic migration operations, the throughput fluctuation of adjacent instances is less than 3\%, and latency jitter is controlled within 5\%. By observing the results produced by CoCoServe instances, we can also conclude that scaling operations can ensure correctness.

\section{Conclusions}
\label{sec:conclusion}
This paper presents a novel scaling mechanism that facilitates automated module replication and migration. Building upon this innovation, we introduce \sys{}, an elastic serving system that effectively addresses two fundamental challenges in the deployment of LLMs: dynamic workload fluctuations and high operational overhead. \sys{} achieves significant cost reductions while maintaining robust performance, providing both service providers and researchers with a fine-grained, scalable solution to fully realize the potential of LLM serving. Our comprehensive evaluation demonstrates that under dynamic RPS conditions, \sys{} reduces operational costs by 46\%, improves latency by 14-75\%, and delivers 1.16$\times$-4$\times$ higher throughput on average compared to state-of-the-art solutions.

\parabf{Software Availability:} The software codes have been open-sourced to \url{https://anonymous.4open.science/r/CoCoServe-B77FSHJKFK/} for research usage.

\bibliographystyle{unsrt}
\bibliography{ref}


\end{document}